\RequirePackage{pgf} 
\documentclass{jaa}
\usepackage{natbib}
\usepackage{authblk}
\usepackage{orcidlink}
\usepackage[switch]{lineno}
\usepackage{hyperref}
\usepackage{babel}

\hypersetup{
    colorlinks=true,
    linkcolor=blue,
    citecolor=blue,
    urlcolor=blue,
    pdfborder={0 0 0}  
}

\usepackage{soul}
\usepackage{graphicx}
\usepackage{booktabs}
\usepackage[normalem]{ulem}
\usepackage{romannum}

\newcommand{\revtwo}[1]{{\color{black} #1}}
\newcommand{\revthree}[1]{{\color{black} #1}}

\graphicspath{{./}{figures/}}

\begin{document}\sloppy

\title{Unraveling the Secrets of the lower Solar Atmosphere: One year of Operation of the Solar Ultraviolet Imaging Telescope (SUIT) on board Aditya-L1}


\author[1]{Rahul Gopalakrishnan\orcidlink{0000-0002-1282-3480}\thanks{Corresponding Author: rahulg.astro@gmail.com}}
\author[3,1]{Soumya Roy\orcidlink{0000-0003-2215-7810}}
\author[1,10]{Deepak Kathait\orcidlink{0009-0001-5450-1754}}
\author[1,2]{Janmejoy Sarkar\orcidlink{0000-0002-8560-318X}}
\author[1]{Nived V. N.\orcidlink{0000-0001-6866-6608}}
\author[1,6]{Durgesh Tripathi\orcidlink{0000-0003-1689-6254}}
\author[1,6]{A. N. Ramaprakash\orcidlink{0000-0001-5707-4965}}
\author[8]{Sami K. Solanki\orcidlink{0000-0002-3418-8449}}
\author[3]{Sreejith Padinhatteeri\orcidlink{0000-0002-7276-4670}}
\author[1]{Mahesh Burse}
\author[1]{Rushikesh Deogaonkar\orcidlink{0009-0000-2781-9276}}
\author[1]{Sakya Sinha}
\author[3]{Adithya H. N.\orcidlink{0009-0002-1177-9948}}
\author[5,6]{K. Sankarasubramanian\orcidlink{0000-0003-1406-4200}}
\author[9,6]{Dipankar Banerjee\orcidlink{0000-0003-4653-6823}}
\author[6,7]{Dibyendu Nandy\orcidlink{0000-0001-5205-2302}}
\author[5]{Srikant Motamarri} 
\author[4]{Amit Purohit}
\author[5]{Rethika T}
\author[4]{Sreenath K R}
\author[4]{Priyanka Upadhyay}
\author[4]{Prapti Mittal}
\author[10]{P. R. Prince}

\affil[1]{Inter-University Centre for Astronomy and Astrophysics, Post Bag 4, Ganeshkhind, Pune 411007, India}
\affil[2]{Tezpur University, Napaam, Tezpur, Assam 784028, India}
\affil[3]{Manipal Centre for Natural Sciences, Manipal Academy of Higher Education, Manipal, Karnataka 576104, India}
\affil[4]{Indian Space Science Data Center (ISSDC), ISTRAC/ISRO, Bengaluru, 560094, India}
\affil[5]{U R Rao Satellite Centre, Old Airport Road Vimanapura Post, Bengaluru - 560017,
Karnataka, India}
\affil[6]{Center of Excellence in Space Sciences India, Indian Institute of Science Education and
Research Kolkata, Mohanpur 741246, West Bengal, India}
\affil[7]{Department of Physical Sciences, Indian Institute of Science Education and Research
Kolkata, Mohanpur 741246, West Bengal, India}
\affil[8]{Max Planck Institute for Solar System Research, Justus-von-Liebig-Weg 3, 37077
\revtwo{G\"{o}ttingen}, Germany}
\affil[9]{Indian Institute of Space Science and Technology Valiamala, Thiruvananthapuram, 695 547, Kerala, India}
\affil[10]{Department of Physics, University College, Research Centre, University of Kerala, Thiruvananthapuram, 695034, India}

\twocolumn[{

\maketitle

\corres{rahulg.astro@gmail.com}
\msinfo{Day Month Year}{Day Month Year}

\begin{abstract}
The Solar Ultraviolet Imaging Telescope (SUIT) is an instrument onboard Aditya–L1, the first solar space observatory of the Indian Space Research Organization (ISRO), India, launched on September 2, 2023. SUIT is designed to image the Sun in the 200–400 nm wavelength band in eight narrowband and three broadband filters. SUIT's science goals start with observing the solar atmosphere and large-scale continuum variations, the physics of solar flares in the NUV region, and many more. The paper elucidates the functioning of the instrument, software packages developed for easier calibration, analysis, and feedback, calibration routines, and the regular maintenance activity SUIT during the first year of its operation. The paper also presents the various operations undergone by, numerous program sequences orchestrated to achieve the science requirements, and highlights some remarkable observations made during the first year of observations with SUIT.
\end{abstract}


}]



\doinum{12.3456/s78910-011-012-3}

\artcitid{\#\#\#\#}
\volnum{000}
\year{2025}
\pgrange{1--}
\setcounter{page}{1}
\lp{1}

\section{Introduction}\label{sec:intro}
The Solar Ultraviolet Imaging Telescope \citep[SUIT;][]{suit_test_calib,suit_main_2} is a solar remote sensing payload onboard the Aditya-L1 \citep{aditya_seetha_megala_2017, aditya} mission of ISRO. The spacecraft was launched on September 02, 2023 from the Satish Dhawan Space Centre (SDSC) in Sriharikota, India. SUIT is designed to record full-disk as well as partial-disk images of the Sun in the near and mid ultraviolet wavelength range of 200{--}400~nm, providing new insights into the dynamics of the photosphere and chromosphere. It has eight narrow-band and three broadband filters to observe various layers of the solar atmosphere in specific wavelength ranges. These observations are crucial not only for the studies of solar atmospheric dynamics but also for monitoring the spatially resolved solar spectral irradiance in the wavelength range central to the photochemistry of ozone and oxygen in Earth's stratosphere. 

The key science goals of SUIT include but are not limited to, the understanding of the dynamic coupling of the lower solar atmosphere, physics of solar flares in NUV, measurement and monitoring of spatially resolved solar spectral irradiance, and dynamics of chromospheric filaments and prominences etc. To address its science goals, SUIT is enabled with various program sequences, allowing users to observe the Sun in different observation modes.


\begin{table*}[h]
\centering
\renewcommand{\arraystretch}{1.3} 
\setlength{\tabcolsep}{8pt} 
\begin{tabular}{lccc}
\toprule
\textbf{Filter ID} & \textbf{Central Wavelength (nm)} & \textbf{FWHM (nm)} & \textbf{Remarks} \\
\midrule
NB01  & 214.0     & 11.0   & Continuum \\
NB02  & 276.7     & 0.4    & Continuum \\
NB03  & 279.6     & 0.4    & Mg~{\sc II}~k \\
NB04  & 280.3     & 0.4    & Mg~{\sc II}~h \\
NB05  & 283.2     & 0.4    & Continuum \\
NB06  & 300.0     & 1.0    & Continuum \\
NB07  & 388.0     & 1.0    & CN Band \\ 
NB08  & 396.85    & 0.1    & Ca~{\sc II}~H \\
BB01  & 200--242  & 42.0   & Herzberg Continuum \\
BB02  & 242--300  & 58.0   & Hartley Band \\
BB03  & 300--360  & 40.0   & Huggins Band \\
\bottomrule
\end{tabular}
\caption{List of science filters available on SUIT. NB and BB stand for narrowband and broadband filters, respectively. \citep{suit_main}}
\label{sc_comb_fil}
\end{table*}

SUIT was switched ON on November 20, 2023, and the first light was taken on December 6, 2023, while the spacecraft was still in the cruise phase. Since then, it has continuously been taking observations, initially for the payload verification and later under the guaranteed science time during the first year of observation. We have established a payload operation centre (POC) for SUIT at IUCAA that liaises with mission operation on day to day basis for planning calibration and science observations, health monitoring etc.

In this paper, we describe the payload operation performed during the first year. In \S\ref{sec:suit} we discuss the technical details of the payload and its observation modes. In \S\ref{sec:commission} we present the major events during the commissioning phase. The calibrations performed for SUIT are presented in \ref{sec:calibration}. A glimpse of the year-long observation is presented in \S\ref{sec:observations}. We describe a day of SUIT observation planning in \S\ref{sec:nominalday}, the data products in \S\ref{sec:data}, and give the concluding remarks in \S\ref{sec:conclusion}.

\section {The SUIT instrument}\label{sec:suit}
SUIT is a two-mirror off-axis Ritchey–Chrétien telescope with a primary mirror of diameter 140.8 mm and a focal length of 3.5 m. It is equipped with eight narrowband and three broadband filters, each serving specific scientific goals \citep[see Table.~\ref{sc_comb_fil} and][for details]{sc_filt}. At the image plane, it has a $\mathrm{4096~\times~4096}$ pixels UV-enhanced back-thinned back-illuminated CCD with a pixel size of $\mathrm{12~\mu m}$. Each pixel corresponds to approximately 502~km on the solar surface when viewed from the L-1 point, providing a field of view (FOV) of $\mathrm{1.5~R_{\odot}}$. 

\begin{figure*}[ht!]
  \centering
  \includegraphics[trim={0cm 4cm 0cm 1cm},clip,width=0.75\textwidth]{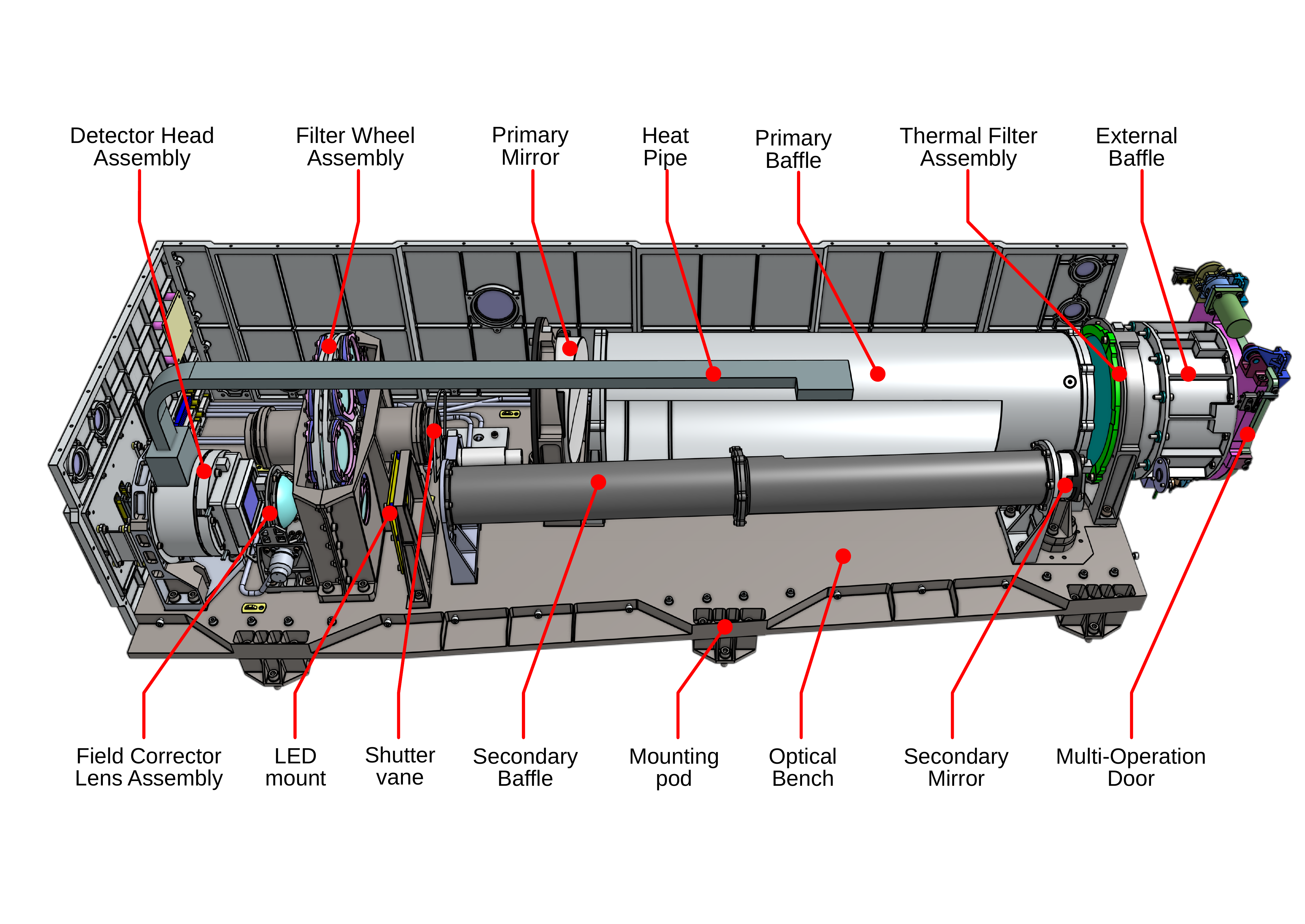}
  \caption{\centering Schematic diagram of the internal components of SUIT.}
  \label{fig:1_suit_diagram} 
\end{figure*}

The SUIT payload architecture comprises three main components: the telescope assembly, payload electronics (PE), and filter wheel electronics (FWE). The telescope assembly includes the multi-operational entrance door \revtwo{(so named because multiple open–close operations are possible)}, a thermal filter that transmits only 0.3\% of the NUV radiation and blocks most of the visible light ($\approx$ 99.75\%), primary and secondary mirrors, a shutter vane mechanism to control exposure time, filter wheel mechanism to select the bandpass of desired wavelength, field corrector lens, and the CCD. The PE unit, located next to the CCD, manages observation modes and flare detection, while the FWE is responsible for the independent movement of the two filter wheels to produce the desired science filter combination. Figure~\ref{fig:1_suit_diagram} shows the schematic diagram of the SUIT assembly depicting the internal configuration. Table \ref{tab:instrument} gives the instrument's critical characteristics.

SUIT is capable of taking full-disk images of the sun with $\mathrm{4096~\times~4096}$ pixel resolution at a cadence of 16 seconds, 2$\times$2 pixels binned ($\mathrm{2048~\times~2048}$) full disk images with a cadence of 8 seconds, and Region of Interest (RoI) images with the fastest possible cadence of 6 seconds in the default size (704$\times$704 pixels) — the read-out time changes with the location of the RoI, causing the cadence to change with it. SUIT has a daily data allowance of 100 gigabits per day. Therefore, to optimize payload output, the default program sequence is designed to take full-disk images in all filters every 2.4 hours, binned full-disk images in the NB03 filter every minute, and RoI images in all narrow band filters with cadences of about a minute. All program sequences were fine-tuned to meet the requirements and data volume restrictions using simulators developed by the SUIT team at IUCAA and tested on the engineering model of the instrument at ISRO before uploading to the spacecraft.

\begin{table}[!ht]
\centering
\resizebox{0.95\linewidth}{!}{%
\begin{tabular}{l r}
\toprule
\textbf{Parameter} 							& \textbf{Value} \\
\midrule
Telescope Design    & Off-axis Ritchey Chr\'{e}tien \\
Wavelength range    & 200 nm - 400 nm.\\
Focal ratio    & f/24.8\\
Bandpass            & 8 Narrow-band, 3 Broad-band \\
Detector            & 4096$\times$4096, back-thinned, back-\\ 
                    & illuminated, UV-enhanced CCD\\
Pixel size          & 12 $\mu$m $\times$ 12 $\mu$m\\
Plate Scale			& $0.7''$/pixel\\
\bottomrule
\end{tabular}}
\caption{\centering SUIT Instrument Specifications \citep{suit_test_calib}}\label{tab:instrument}      
\end{table}

SUIT currently has 26 program sequences stored onboard. These sequences include parameters of exposure, image type, mode of observation, science filter selection, etc. Each program sequence is designed to meet specific science requirements. The program sequences also manage the flare mode. SUIT automatically switches to this observation mode when the onboard intelligence detects a solar flare. SUIT obtains external flare triggers from other instruments onboard Aditya-L1, such as the Solar Low Energy Spectrometer (SoLEXS) and the High Energy L1 Orbiting X-ray Spectrometer (HEL1OS). Additionally, SUIT also has a self-flare detection module, which uses the binned NB03 images to check for a sudden jump in counts with a steady rise. The instrument localizes the flare on the Sun disk and takes RoI observations of the flare at high cadence \citep[see][for details]{suit_algo}. For more information on the onboard intelligence and flare trigger module, refer \citep[]{suit_main_2, suit_algo}. Additionally, the scientific community can submit proposals with their requirements and goals, and the SUIT team can generate and upload new program sequences to the payload after consultation with mission operations.

\section{Commissioning of the payload}\label{sec:commission}

\begin{figure*}[ht!]
  \centering
  \includegraphics[trim={0cm 0cm 0cm 0cm},clip,width=0.95\textwidth]{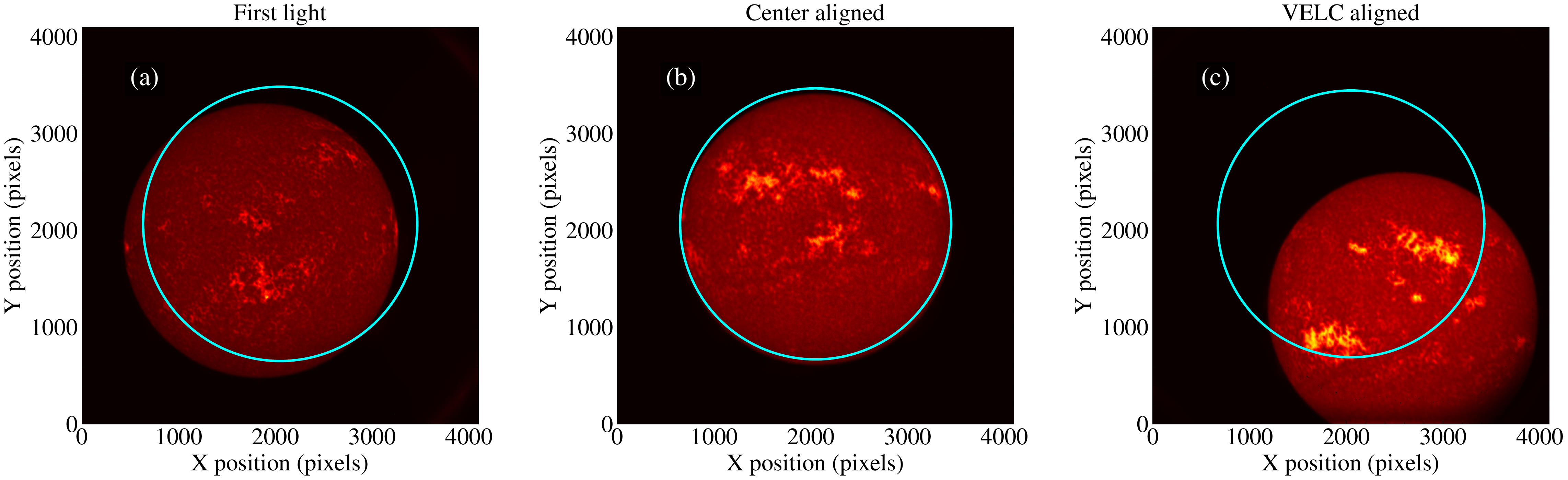}
  \caption{The three positions of the Sun on the CCD in SUIT's one year of operation are shown here. Panel (a) shows the position of the sun in the first light image. Panel (b) shows the position when the Sun is placed on the CCD center, and Panel (c) is the final position after aligning the spacecraft to occult the Sun's disk for the VELC instrument. The cyan circle represents the ideal position of the Sun at the CCD center in all panels. }
\label{fig:sun_pos} 
\end{figure*}

One of the first objectives after the power-on is payload commissioning. For that purpose, we recorded calibration images, such as dark, bias, and engineering mode LED images, to check the instrument's overall functioning. Moreover, we could define the dark current and set the bias limits using these calibration images. After a detailed study of these calibration images, we defined the bias limit to 500 counts. These observations also demonstrated that the CCD produces a dark current of $1.5~\mathrm{e^-.pix^{-1}.s^{-1}}$ at a CCD temperature of $-50^{\circ}C$. \revtwo{The signal strength ranges between 15000 and 36000 electrons per pixel, in the quiet-sun, for observations in various filters with varying exposure times. The dark current at the longest viable exposure time of 2 seconds, along with a readout time of 18 seconds, is 30 electrons per pixel ($1.5 e^{-1} px^{-1} s^{-1}$). The noise associated with this dark current is $\approx 5-6$ electrons.}

\revtwo{On December 6, 2023, we obtained the first full-disk images across all filters while the spacecraft was in its cruise phase}. We noticed that on the door opening, the Sun's image was not exactly at the CCD centre as expected, but it was shifted by [-2.31$^\prime$, -1.79$^\prime$] from the centre of the CCD as depicted in the first panel of Fig~\ref{fig:sun_pos}. To re-point the satellite such that the Sun falls at the centre of the SUIT CCD, as in the middle panel of Fig~\ref{fig:sun_pos}, the mission control applied rotational biases about satellite pitch and roll axes.

\begin{figure}[ht!]
  \centering
  \includegraphics[trim={0cm 0cm 0cm 0cm},clip,width=0.45\textwidth]{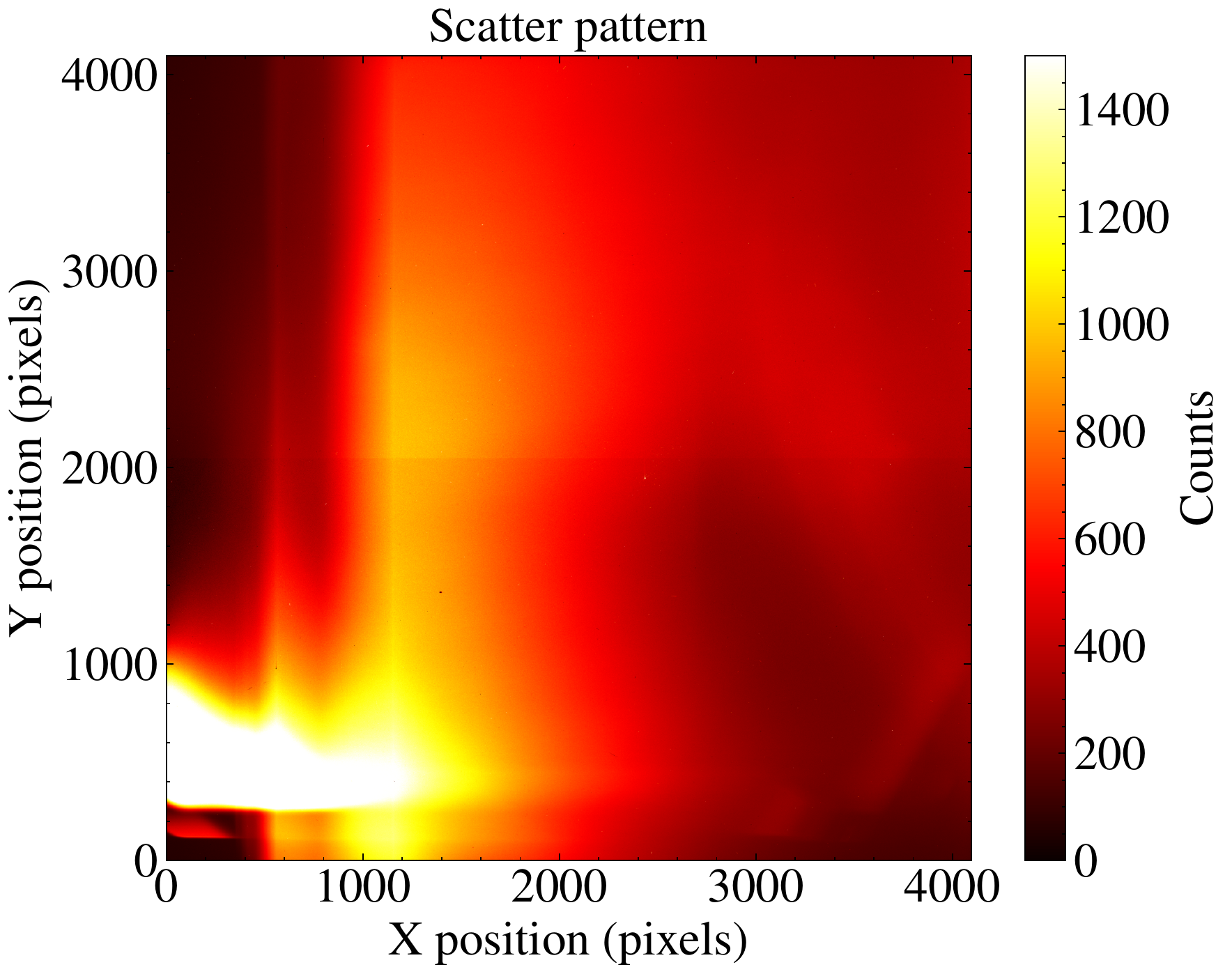}
  \caption{The first pattern of scattered light seen in the BB03 filter on December 6, 2023. The image was recorded with the entrance aperture door open, while the vane shutter was closed. \revtwo{The CCD readout direction is such that lines above row 2048 are read upward, while lines below are read downward.}}
  \label{fig:scatter_pattern} 
\end{figure}

The initial door-open dark and bias images revealed a pattern of scattered light on the CCD as shown in Fig~\ref{fig:scatter_pattern}. Initially, the calibration plan included capturing two dark and two bias images daily before science observations for bias correction. However, acquiring these calibration images was discontinued as the contribution due to scatter light was two orders of magnitude higher than the dark current values. \revtwo{For reliable dark images, the SUIT door may be closed. However, the door mechanism has a limited lifetime, which is defined by the number of opening and closing operations. Therefore, the dark images are recorded only when the door is closed for satellite-related activities like station-keeping maneuvers and momentum dumping of reaction wheels}. We now perform the bias correction using the under-scan pixel columns on either side of the final image \citep[refer][for more details]{det_character}. We determine the bias values for each rows by taking the median of 44 pixels, excluding three pixels from each edge to avoid potential bleeding effects. The bias baseline was fine-tuned for each quadrant and was set to \revtwo{466 counts} \citep[see][for further details]{suit_test_calib}.

As is apparent from the scatter image shown in Fig.~\ref{fig:scatter_pattern},  the amount of scattered light is not the same throughout the CCD. Therefore, depending on the location of the RoI, the scatter pattern is different. Moreover, it is different for the two different starting positions of the rotating shutter vane.
Considering all these factors, scatter models have been created for each shutter start position and each filter combination \citep{suit_test_calib}. Additionally, a separate algorithm is developed to mimic the readout process of the RoI region to create readout scatter patterns. These models are used for scatter correction in the Level-0 to Level-1 pipeline.

The spacecraft was finally inserted into a halo orbit around the Sun-Earth L1 point on January 6, 2024, following which the Visible Emission Line Coronagraph (VELC), another payload on Aditya-L1, was switched on. To produce coronagraphic observations with VELC, the spacecraft has to be aligned so the occultor fully blocks the solar disk. Under this configuration, the solar disk was centred at approximately [1216 pix, 2587 pix] on the SUIT CCD, causing a significant portion of the solar southern hemisphere to fall out of the SUIT field of view, as shown in the panel (c) of Fig. \ref{fig:sun_pos}.

Our observations suggest that the scatter pattern on the \revtwo{solar image} depends on the way the spacecraft is aligned to the Sun. Therefore, we have to develop two scatter models {--}, one when the Sun was located at the SUIT CCD centre (i.e., SUIT aligned position) and another for VELC aligned position.

\section{Software developed for payload and mission operation}\label{sec:software}
To aid the commissioning of SUIT and mission operation, we developed various tools.
\subsection{Quick Look Display}
The Quick Look Display (QLD) software is developed at the SUIT POC to process raw data by converting it into FITS format and displaying the corresponding images. It was initially deployed at ISRO Telemetry Tracking and Command Network (ISTRAC) and the Indian Space Science Data Centre (ISSDC) during the early operations phase to enable quick analysis of recently downloaded data. The software presents the most recent images across all 11 filters in full-disk mode, 8 in ROI (Region of Interest) mode, and 2$\times$2 binned images. As data is received, QLD serves as a first-level data visualization and quality check tool, helping identify anomalies such as issues with program sequences, command uploads, and spacecraft pointing errors. \revtwo{The modified version of the QLD software is now publicly accessible through the ``Latest Observations'' section on the SUIT website\footnote{SUIT latest observation page: \href{https://suit.iucaa.in/suitdaily}{https://suit.iucaa.in/suitdaily}}.}

\subsection{Sun-center finder}
We developed the \textit{Sun Center Finder Toolkit} to estimate the Sun's geometric centre from the SUIT images and made it operational immediately after the SUIT door was opened. This toolkit helps provide spacecraft-pointing feedback to the \revtwo{mission team} using the recorded full disk images in synoptic mode.

The algorithm analyses intensity profiles along thin vertical and horizontal strips of the image. It detects the solar limb based on sudden changes in intensity near the edge of the Sun. A pair of moving kernels is used to identify the limb location by comparing local intensity variations. Once all limb points are collected, a curve-fitting routine determines the Sun's centre and radius in pixel units \citep[for further details refer to][]{pipeline_paper}. This information is used to track the spacecraft's drift over time. During manoeuvres, the sun centre coordinates are regularly sent to the mission team to ensure accurate pointing. \revtwo{The Sun’s center position exhibits a periodic variation with a period of approximately 2 hours and a peak-to-peak amplitude of about $20^{''}$ in the x-direction and $30^{''}$ in the y-direction.}

    
While the algorithm achieves sub-pixel precision during the SUIT-aligned scenario, its accuracy decreases in the VELC-aligned scenario due to distortion and vignetting, leading to a standard error of about 3 pixels.


\section{Calibration routine and payload maintenance}\label{sec:calibration}
During the first year of operation, we performed various calibration runs for SUIT. This included dark and bias images with the entrance aperture door closed, LED calibration images, and solar observations by varying the spacecraft pointing, such that the Sun appears at different positions on the CCD. The later observations were used to create the CCD master flat image.

\begin{itemize}
 \item The Aditya-L1 spacecraft performs a momentum dump of its reaction wheels every month. During this phase, the SUIT aperture door is automatically closed, and the spacecraft thrusters are fired. We take this opportunity to record door-closed dark and bias images and LED illumination images, as they are devoid of scattered light from the Sun. \revtwo{SUIT has 16 LEDs. 8 LEDs operate at 258 nm, and the other 8 operate at 356 nm.} 
    
\item From January 28, 2024, to February 15, 2024, the SUIT recorded special observations for flat-field calibration. For this purpose, we used custom-built program sequences where the spacecraft was pointed such that the Sun illuminates different portions of the CCD to record full-disk images in all filters. We also recorded dark, bias, and LED illuminated images at each pointing of the spacecraft. While the LED images were used to compute the pixel response non-uniformity (PRNU) in each pixel, the full-disk Sun images at different locations on the CCD were used to determine the scatter light pattern and large-scale flat field calibration images \cite{suit_test_calib}.
    
\item To perform absolute calibration, we observed Sirius-A, a stable star with a known variability limit in the NUV regime, from November 28 to December 15 2024. This data is currently being analysed for PSF determination, photometric calibration, and understanding the pointing accuracy of the spacecraft.
    
\item The flare threshold finalisation is one of the main steps in calibrating the flare detection module. As presented in \S\ref{sec:suit}, SUIT can localise a flare if the counts observed are above a certain threshold. This has to be fine-tuned for quick localisation and to avoid false detections \cite{suit_algo}. For this, we simulated the onboard algorithm on the ground with HEL1OS data and then determined and uploaded the threshold. We completed this process on March 15, 2024. We further note that due to throughput variation of the CCD, this threshold needs continuous updating.

As alluded to earlier, the amount of scattered light differs for two diametrically opposite openings of the rotating shutter vane. Therefore, we adjusted all the program sequences to record $2\times2$ binned images in one particular shutter position so that a single threshold value may be used for the flare trigger.

\item We noticed a degradation in throughput by March 2024. Based on the analysis of the recorded images, we concluded that such degradation is due to the deposition of contaminants on the CCD, which is maintained at $-50^{\circ}C$ under nominal conditions, making it the coldest component inside the payload cavity. To eliminate these contaminants, we bake the detector assembly to $\approx 28^{\circ}C - 30^{\circ}C$. The first baking was performed between April 23 and May 14, 2024.
Observations below 300 nm were seen to be the most affected by the contaminants. However, a significant improvement in throughput was also noticed in these shorter wavelengths. The throughput was seen to improve by $> 4$ times in the LED illuminated images of 255 nm at the center of the detector. However, a change by $\approx 1.1$ times was seen for the 355 nm LED illuminated images at the center of the detector.
We have also tested to see if baking the full payload is better for eliminating the contaminants. A complete payload baking was performed between August 1 and September 10, 2024. The comparative analysis of CCD baking vs. complete payload baking is underway. Any conclusive results may be understood after the next cycle of baking. Note that no solar observations can be carried out during baking. We have now put in a process for a regular baking cycle, which will be carried out approximately every three to four months. 
\end{itemize}
    
The test and calibration of SUIT led to several changes in the payload program sequences. The finalized sequences were uploaded on June 14, 2024. The exposure values were changed for each filter based on the counts measured in the images on May 24, 2024. For further details on the various in-orbit test and calibration, please refer to \cite{suit_test_calib}. 

\revtwo{
A comparison between the design requirements and the results from the in-orbit calibration tests is presented in Table~\ref{tab:calib_tests}. Some of the tests are still ongoing. The table summarizes the key performance parameters of SUIT, including the field of view, point spread function (PSF), plate scale, modulation transfer function (MTF), detector characteristics, and photometric accuracy. The results indicate that most of the measured parameters meet or exceed the design specifications, which confirms the stable performance of the instrument in orbit.}

\revthree{With respect to the science goals evaluated against these performance parameters, the in-orbit results show that SUIT largely satisfies the expected spatial and temporal requirements. For irradiance-related studies (long-term solar irradiance, microflare enhancement, and large-flare spectral energy distribution), the required spatial coverage ranges from $\approx10''$ to $>50''$, with time scales varying from seconds and minutes to hours and even mission lifetime. For studies of dynamic solar features such as jets, flares, waves, and prominences, the required spatial range is $2''– 100''$ with temporal evolution on minute-to-hour timescales.

The achieved performance meets the spatial and temporal requirements for all these cases except for the smallest spatial scale in the kinematic studies, where $3''$ is achieved instead of the required $2''$. All temporal requirements, however, are fully satisfied. In addition, for long-term irradiance measurements, the solar disk must be at the center of the SUIT CCD to measure the total solar irradiance. Overall, the instrument performance in orbit supports the relevant science objectives within the expected operational limits.}

\begin{table*}[h!]
\centering

\renewcommand{\arraystretch}{1.3}
\begin{tabular}{lll}
\hline
\textbf{Parameter} & \textbf{Specification} & \textbf{Onboard Test Result} \\
\hline
Field of View & $0.39^\circ$ radius & $0.39^\circ$ radius \\
PSF (80\% EE) & $1.4''$ & To be tested \\
Plate Scale & $0.7''$/px & $0.69''$/px \\
MTF & 10\% at 42 lp/mm & To be tested  \\
Read Noise & $< 10~e^{-}$ & $9~e^{-}$ \\
Bias & $< 500$ ADU & $466$ ADU $\pm 0.12\%$ \\
Mean Dark Signal & $< 10~e^{-}$/px/s & $1.5~e^{-}$/px/s \\
PRNU (Post Correction) & $< 1\%$ & $\sim 0.33\%$ at 255 nm \\
& & $\sim 0.43\%$ at 355~nm\\
Flat Field Uniformity & $< 1\%$ & $< 0.11\%$ (Post Correction) \\
Total Photometric Error & $< 1\%$ & $0.56\%$ (Post Correction) \\
\hline
\end{tabular}
\caption{
\revtwo{Summary of the main calibration and performance verification tests conducted for SUIT. The table compares the instrument specifications defined during design with the corresponding results measured during in-orbit operations.}
}

\label{tab:calib_tests}
\end{table*}

\section {Early years of observations}\label{sec:observations}
The SUIT instrument has operated through various phases and has provided crucial data for solar studies, including many interesting flares and eruptions of the Sun. Observations from December 2023 to April 2025 are referred to in this section. The default synoptic mode has been the major contributor to the data taken by SUIT, generating around 100 gigabits of data daily. The default synoptic mode takes around 10,000 images per day - about 200 images of full-frame, 1440 images (every minute) of binned, and the rest of them being RoI images. These numbers can vary depending on the observation duration planned and the flare-trigger mode activation. We also recorded observations in engineering mode for both calibration and science purposes. Note that most of the data are taken with the spacecraft prioritizing the VELC alignment with the Sun, which has restricted SUIT to only set ROI locations on the north-east limb.

\subsection{Automated Flare observations}
\begin{figure*}[ht!]
\centering
\includegraphics[trim={0cm 0.31cm 0.1cm 0cm}, clip, width=0.95\textwidth]{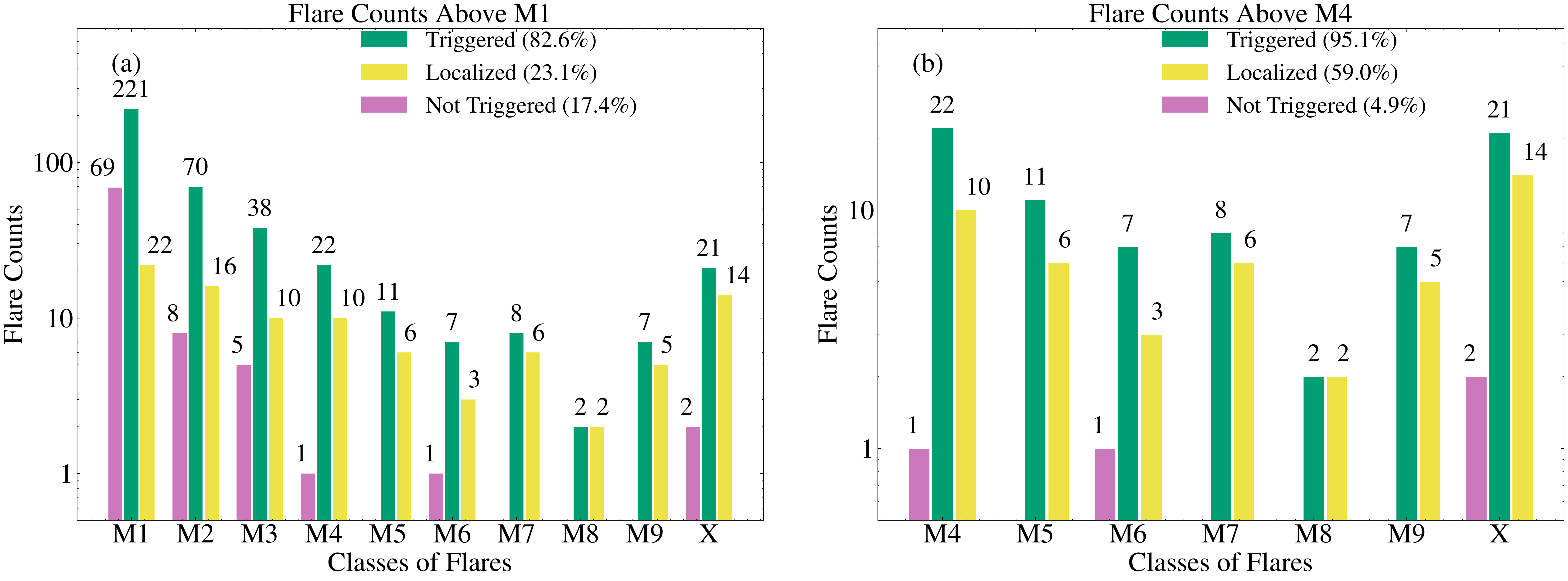}
\caption{\textit{Panel (a)}: Flares above GOES Class M1 observed by SUIT from March 13, 2024, to February 25, 2025, the distribution of triggered flares that were successfully triggered (green bars), localized (yellow bars) and not triggered (magenta bars) across various GOES classes. \textit{Panel (b)}: Same as panel (a), but restricted to flares above GOES Class M4.}
\label{fig:flare_combined}
\end{figure*}

The onboard flare detection and localization algorithm of SUIT is designed to autonomously identify flares of M1-class and above and automatically switch to RoI observation mode to observe solar flares. This is is a two-step process {--} generating a flare trigger flag suggesting the start of a solar flare anywhere on the solar disk, followed by localizing the flare on the solar disk so as to perform RoI observations defined around the localized flare.

\subsubsection{Self trigger and external trigger:} There are three independent ways to generate a flare trigger flag, i.e., the self trigger flare flag generated using SUIT data and the external trigger flags generated using HEL1OS and that obtained from SoLEXS measurements. 

The self-trigger mechanism uses the 2k$\times$2k binned images recorded every minute and checks for local brightenings in super pixels of the size 32$\times$32 in continuous three frames, indicating the initiation of a flare. In the external trigger, while SoLEXS directly provides triggers as binary flags, HEL1OS sends a data stream to SUIT payload where we perform a real-time calculation to generate the flare flag, based on the algorithm \citep[see][for details of the algorithm]{suit_algo}.

The trigger algorithm for both self and external trigger rely on three fixed thresholds to confirm the flare onset, which are fine-tuned to reduce false positives. We note that both the external trigger and self-trigger algorithms are optimized for detecting larger flares (M1 and above), but they face limitations when dealing with lower-intensity flares.

\subsubsection{Localization:}
When a flare is detected and a trigger is raised, the flare localization module gets into action to to locate the flare on the solar disk by stopping all the other activities onboard. In this module, SUIT records four continuous Mg~II h binned images with a cadence of $\approx$10-12~s and runs the onboard localization algorithm \citep[see][for further details]{suit_algo}. Once a flare is localized, the payload goes into a 2-hour flare mode, with SUIT imaging an RoI of 704$\times$704 pixels around the localized flaring region.

During this 2-hour time window, all other triggers are ignored on board, and the flaring region is imaged at high cadence with automated RoI tracking and exposure-time adjustment. These activities are automatically performed onboard, without any ground intervention.

Note that the flare localization module is currently configured to run for up to a maximum of 10 minutes to locate a flare. If a flare is localized within this 10 minutes window, the payload goes into flare mode. Otherwise, it continues the regular operation where it had left. The localization duration can be changed to a custom value. We further note that, based on the requirement of an observational proposal, the flare flags may completely be ignored.

\subsubsection{One year observation of flares:} We present here a statistics highlighting the efficiency of the trigger and localization module using the observations recorded during 13 March 2024 (the date on which all SUIT trigger modules were deployed) and 25 February 2025 (the latest baking start date of the SUIT CCD). During this time window, there were 838 solar flare events of class M1.0+, obtained from from the Heliophysics Event Knowledge base \citep[HEK;][]{hurlburt_heliophysics_2021}. Among these, 493 events occurred during the synoptic operation of SUIT, where flare trigger and localization modules were active. All these events were observed in the 2$\times$2 binned Mg~\Romannum{2}~k observation with a cadence of $\sim$1~min.

The onboard flare trigger algorithm (either external or self-trigger) successfully triggered 407 flares of the aforementioned 493 flares, demonstrating a trigger efficiency of approximately 82.6\%. Among the 86 flares that were not triggered, 18 occurred while the instrument was operating in flare mode -- during which trigger generation is disabled by design.

Among the triggered 407 flares, 94 ($\sim$ 23.1\%) were successfully localized by the onboard mechanism. In Fig.~\ref{fig:flare_combined}.a, we illustrate the fraction of triggered (green), not triggered (magenta), and localized (yellow) flares in the bar diagram according to their respective GOES class. As it appears, while the trigger is generated with and efficiency of $\approx$83\%, the localization is only about 23\% of the triggered flares. While manually inspecting data on the ground, we find that most of the non-localized flares are of lower energy, i.e., below class M4. In Fig.~\ref{fig:flare_combined}.b, we plot the statistics of flares of class M4+. We note that for such flares, the improvement in the trigger as well as localization efficiency is tremendous. During the aforementioned time, there were 82 M4+ flares, out of which 78 were successfully triggered, giving an efficiency of $\approx$95.1\%.

Among these triggers, 46 were successfully localized on board. This gives a $\sim$ 59.0\% localization efficiency, which is drastically better than the localization efficiency for M1+ flares ($\sim$ 23.1\%).

\begin{figure}[ht!]
    \centering
    \includegraphics[trim={3cm 1.2cm 1.8cm 0.8cm}, clip, width=0.9\linewidth]{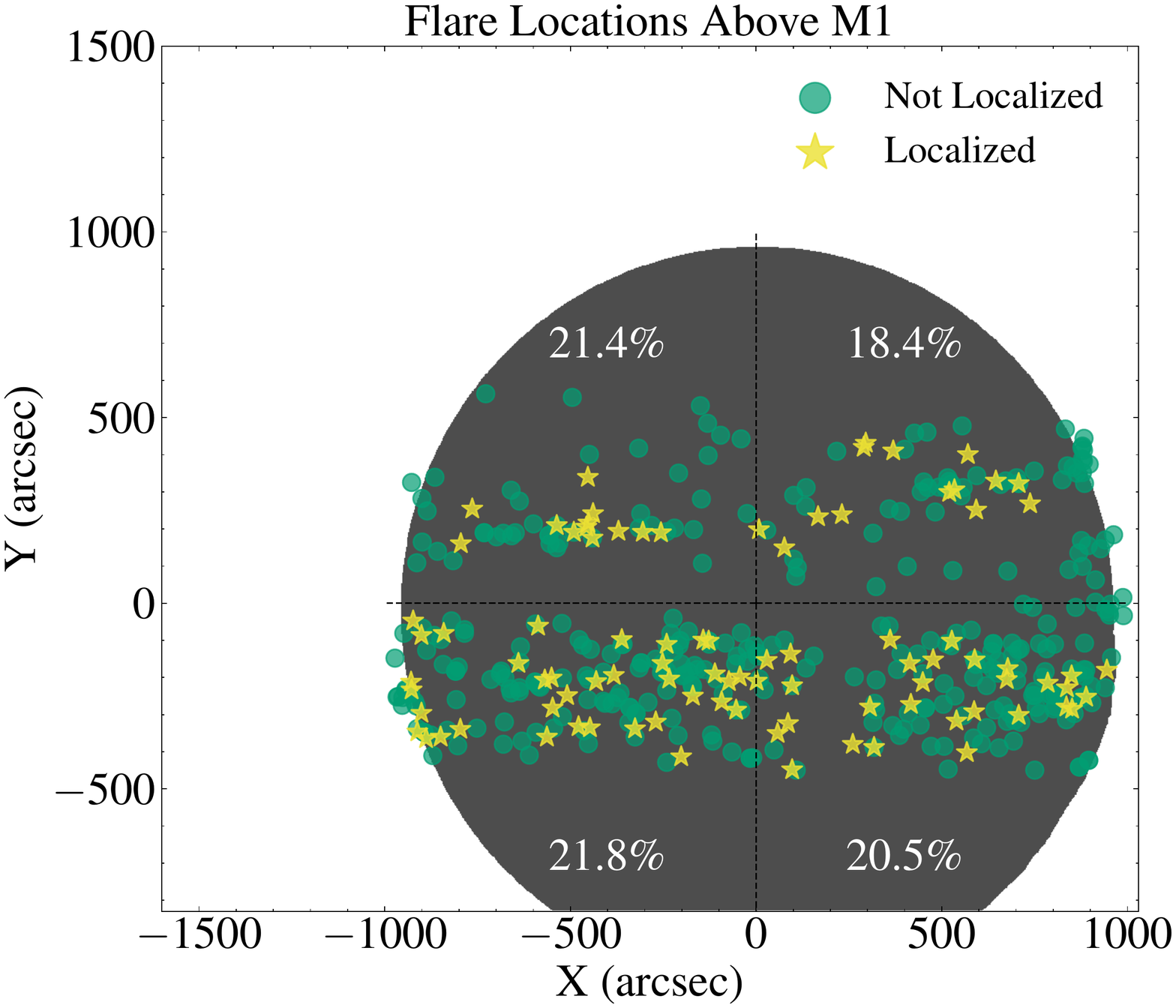}
    \caption{Locations of successfully localized M1+ flares (yellow stars) and those not localized (green circles).}
    \label{fig:flare_loc}
\end{figure}

In Figure \ref{fig:flare_loc}, we also present the locations of flares localized (yellow stars) and not localized (green circles) during the aforementioned time period and for flares M1+. In each quadrant, we also mark the localization efficiency. It is also worth mentioning here that the locations of all the flares exhibit more flares in the southern hemisphere of the Sun, alluding to higher southern activity in this solar cycle. This has been predicted for the current cycle from models of hemispheric sunspot number predictions \citep{rodriguez24}.

\subsection{Coordinated observations}

\begin{table}[h!]
\centering
\begin{tabular}{@{}l l@{}}
\hline
\textbf{Period} & \textbf{Observatory} \\
\hline
10--16 Jul 2024 & Sunrise~III \\
28--30 Mar 2024 & Solar~Orbiter \\
22 May 2024 -- & Dunn~Solar~Telescope \\
22 May 2024 -- & MAST \\
Dec 2023 -- Jul 2025 & IRIS: SUIT calibration\\
& observations \\
Aug 2025 -- & IRIS coordinated observations \\
\hline
\end{tabular}
\caption{\revtwo{Summary of coordinated observation campaigns conducted with SUIT in collaboration with various observatories.}}
\label{coordinated-obs}
\end{table}

In addition to its regular observations, SUIT has participated in joint observation campaigns with other instruments. The joint observation with Solar Orbiter was conducted on March 28, 2024, and the observation campaign with the balloon-borne Sunrise observatory took place on April 15, 2024. During these joint observations, full-disk images were taken in all filters, with the spacecraft pointed to keep the Sun at the center of the CCD. 
    
Another notable attempt was to observe comet C/2024 S1 (ATLAS) during its perihelion on October 28, 2024 at a distance of about 0.008 AU from the barycenter of the Solar System. However, the comet disintegrated, thereby making it undetectable with SUIT.
    
Additionally, the Interface Region Imaging Spectrograph \citep[IRIS;][]{iris} regularly records data for coordinated observations with SUIT. Ground-based telescopes like the Multi Application Solar Telescope (MAST) at the Udaipur Solar Observatory and the Dunn Solar Telescope also have regular coordinated observations with SUIT. \revtwo{A calendar overview of the coordinated observations is given in Table~\ref{coordinated-obs}.}
    
\subsection{Some interesting events}\label{sec:events}
Here, we present a selection of particularly interesting events captured by SUIT, highlighting their key observational characteristics and scientific significance. By comparing these events with simultaneous observations from other instruments, we aim to gain a comprehensive understanding of their underlying physical processes and their broader impact on solar atmospheric dynamics. In Table.~\ref{events}, we highlight the details of the featured observations.
    
    \begin{table}
    \centering
    \begin{tabular}{ccc}
    \toprule
    \textbf{Observation} & \textbf{Location} & \textbf{Description} \\
    \textbf{date} & \textbf{(arcsec)} & \textbf{of the event} \\
    \midrule
    Dec 6, 2023 & [$-900''$, $0''$] & Off-limb tornado\\
    Dec 31, 2023 & [$-950''$, $100''$] & Off-limb flare \\ 
    & & ejected plasma blob \\
    Feb 22, 2024 & [$-400''$,$400''$] & On-disk flare \\
    May 27, 2024 & [$-900''$,$-350''$] & Off-limb prominence \\
     & & \& flare loops \\
    \bottomrule
    \end{tabular}%
    \caption{Overview of some of the events described in Section~\ref{sec:events}. The `Location' column denotes the solar latitude and longitude of the region in arcseconds.} \label{events}
    \end{table}

    \begin{itemize}
        \item One of the very first interesting observations that SUIT made was a `tornado' visible on the east limb on December 6, 2023.
        A solar `tornado' generally refers to a magnetized rotating plasma structure observed in the solar atmosphere, particularly in the chromosphere and corona \citep{trace, su_solar_2012, wedemeyer_are_2013}.
        These structures are associated with vortical motions extending from the chromosphere to the corona, and are often linked to various dynamic solar phenomena.
        The observation taken in the NB04 (Mg~\Romannum{2}~h) is shown in Fig.~\ref{fig:events1}. On the leftmost panel, the location of the event is shown on a full disk observation with a white dashed box.
        The middle panel shows the region marked by the red dashed box.
        The `tornado' is only visible after saturating the disk (right panel).
        This eruption was associated with a slow (350 km/s) partial halo CME observed by LASCO coronagraphs on the east limb of the Sun.
        
        \item The other interesting event observed during the cruise phase to L1, was an X5 flare on December 31, 2023 at 21:36:00 UTC from the active region NOAA 13536 at the east limb.
        This was the first X-class flare observed by the payload. Since the spacecraft was still in the cruise phase, the onboard flare detection algorithm \citep{suit_algo} was not activated.
        So, we have only (2k~$\times$~2k) observations in NB04 filter (Mg~\Romannum{2}~h). The flare was associated with an ejected plasma blob accelerated at very high velocities (in excess of $\mathrm{\approx~1500~km/s}$).
        The simultaneous hard X-ray (HXR) observations confirm the simultaneous reconnection going on during the acceleration phase of the plasma blob.
        The observation of the flare is shown in Fig.~\ref{fig:events2}. 
        In the left panel, we show a full disk observation with the white box marking the event's position on the disk.
        In the right grid, we show a sequence of observations of the ejection of the plasma blob. For further details on the event and subsequent analysis, refer to \cite{dec_31st}.
        
        \begin{figure*}[ht!]
            \centering
            \includegraphics[trim={0cm 3cm 0cm 3cm}, clip, width=0.9\linewidth]{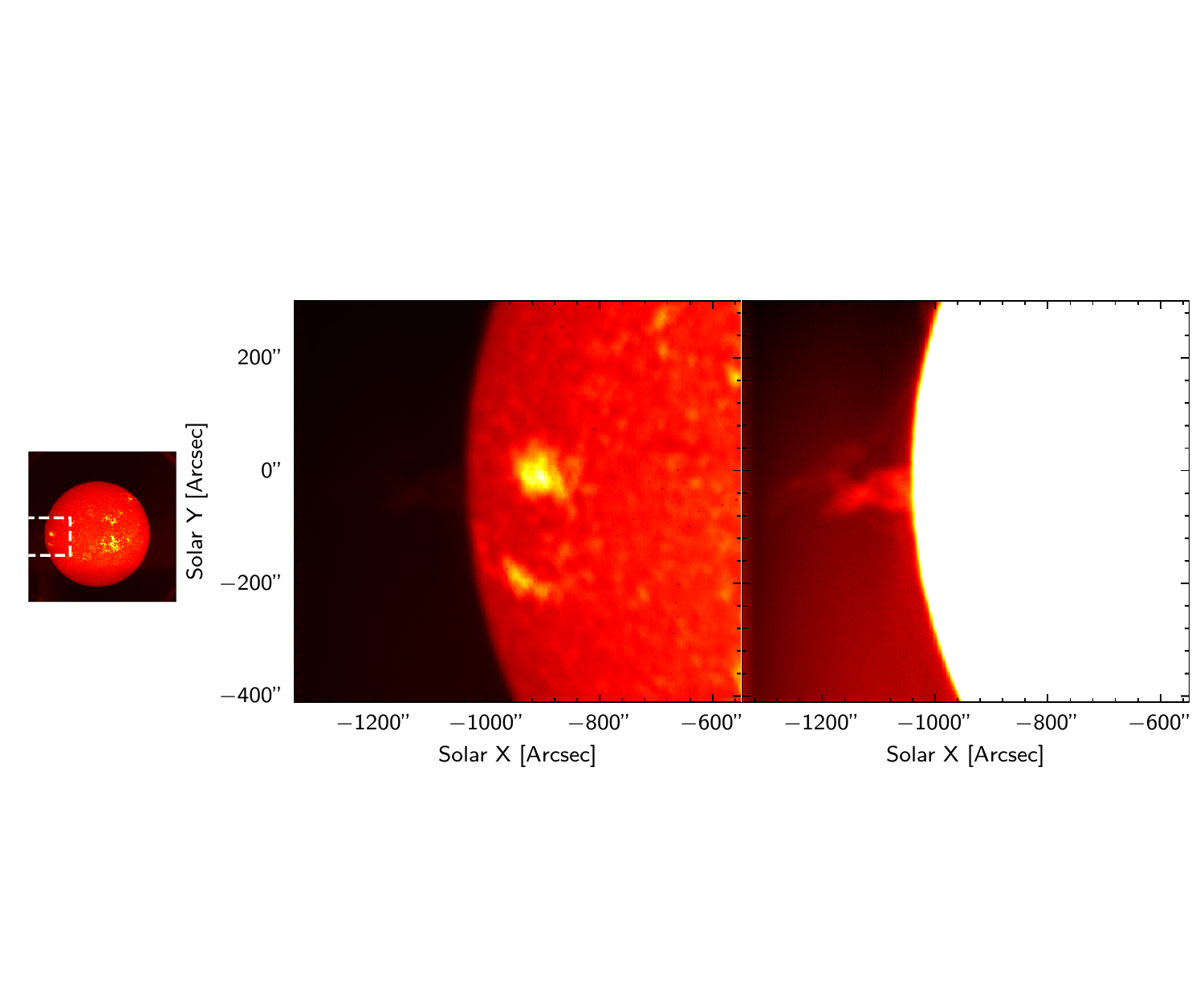} \\ 
            \caption{A solar `tornado' observed on the east limb during the cruise phase on December 6, 2023.}
            \label{fig:events1}
        \end{figure*}

        \begin{figure*}[ht!]
        \centering
          \includegraphics[trim={0cm 0cm 0cm 0cm}, clip, width=0.9\linewidth]{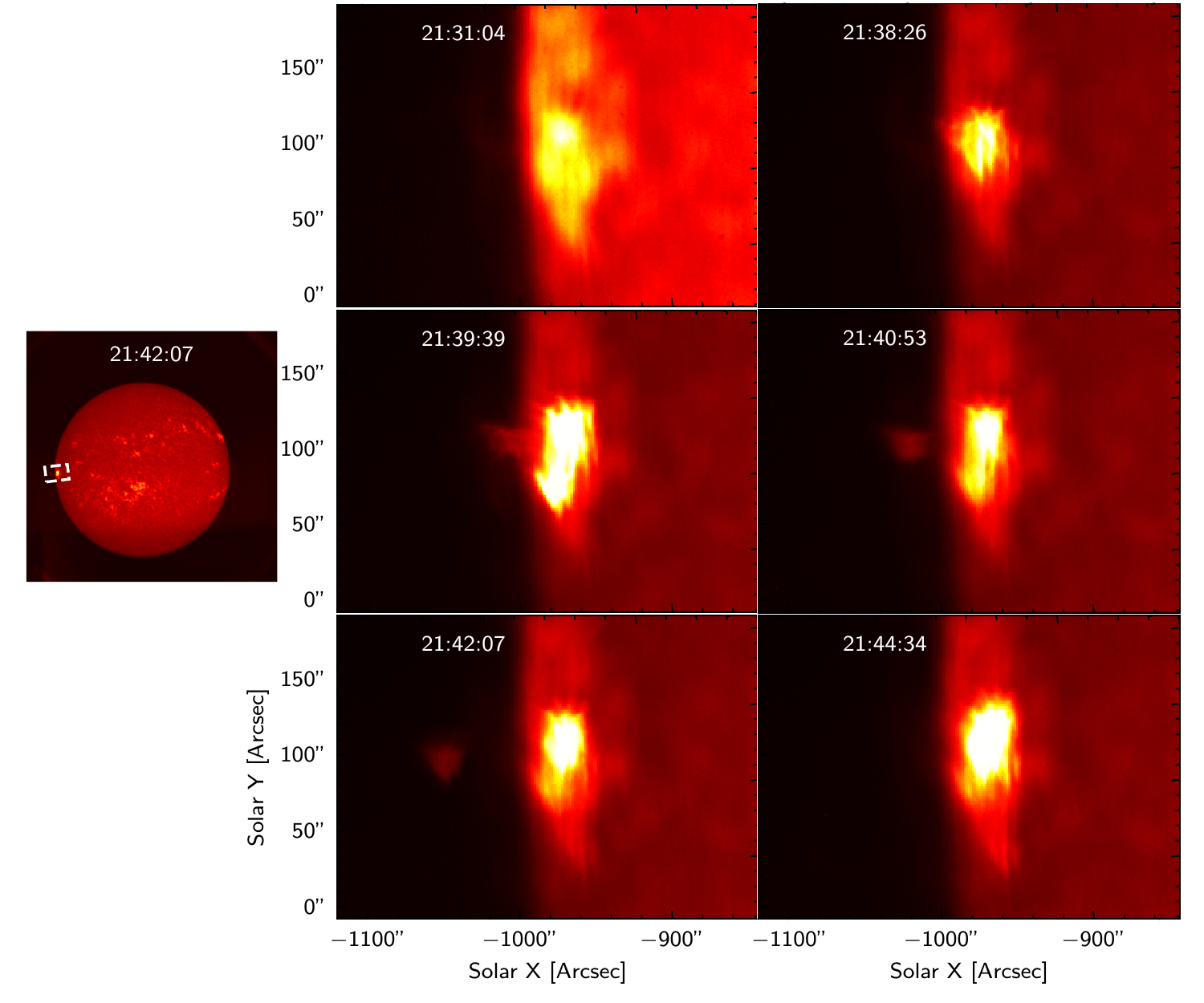}
          \caption{Solar flare and the associated plasma blob ejection were observed on the east limb on December 31, 2023.}
          \label{fig:events2}
        \end{figure*}
        
        \item One of the strongest flares of this solar cycle, an X6.3 flare, happened on February 22, 2024 from the active region NOAA 13590.
        It had a $\beta\gamma\delta$ configuration and produced three flares.
        SUIT missed the first flare, as it was off-pointed to test the stellar calibration program sequences \S\ref{sec:calibration}. SUIT observed the third flare along with other payloads on Aditya{--}L1.
        This was the first reported spatially resolved observation of a solar flare in the passbands centered at 214 (NB01), 276.7 (NB02), 300 (NB06) and 388 (NB07).
        We computed the flare light curve by adding the intensity observed in the 60\% intensity contour of NB03 (Mg~\Romannum{2}~k, 279.6~nm) for the various filters and observations from AIA~1600, 1700~{\AA}, GONG~H{$\mathrm{\alpha}$}, Hard X-ray (HXR) observations from STIX and Soft X-ray (SXR) observations from SoLEXS.
        We observed that all the SUIT line channels i.e. NB03 (Mg~\Romannum{2}~k, 279.6~nm), NB04 (Mg~\Romannum{2}~h, 280.3~nm) and NB08 (Ca~\Romannum{2}~H, 396.85~nm) peak at the same time with STIX HXR, AIA~1600, 1700~{\AA}, GONG~H{$\mathrm{\alpha}$}, NB01 (Herzberg continuum) and the temperature obtained from fitting the SoLEXS spectra.
        In Fig.~\ref{fig:events3} panel (a), we show the SUIT observation of the flare. In Fig.~\ref{fig:events3} panels (A-H), we show the RoI observations of the flare in all eight narrow bands during their respective peaks.
        The observed region is marked with a white dashed box in panel (a). The observation of penumbral bright kernels in the red (NB05) and blue (NB02) wings of Mg {\sc II} are particularly curious.
        Similar observations have been made several times in the past \citep{heinzel14,kleint16,kleint17,kowalski19}. A similar enhancement on the blue side of the Mg~\Romannum{2}~k line was observed in $\approx~\mathrm{2782.56~{-}~2795~\AA}$ by \cite{reetika21}.
        The enhancements observed by SUIT are in the far blue continuum of the Mg~\Romannum{2}~k line $\approx~\mathrm{2767~\AA}$.
        
        These observations are the first spatially resolved observations of a solar flare in these wavelengths.
        This opens up new avenues for exploring the response of the local plasma in these wavelength ranges with coordinated spectroscopic observations and simulations. For further details on the data analysis and conclusions, see \cite{feb_22nd}.
        
        \begin{figure*}
            \centering
            \includegraphics[trim={0cm 11cm 0.5cm 16cm}, clip, width=\linewidth]{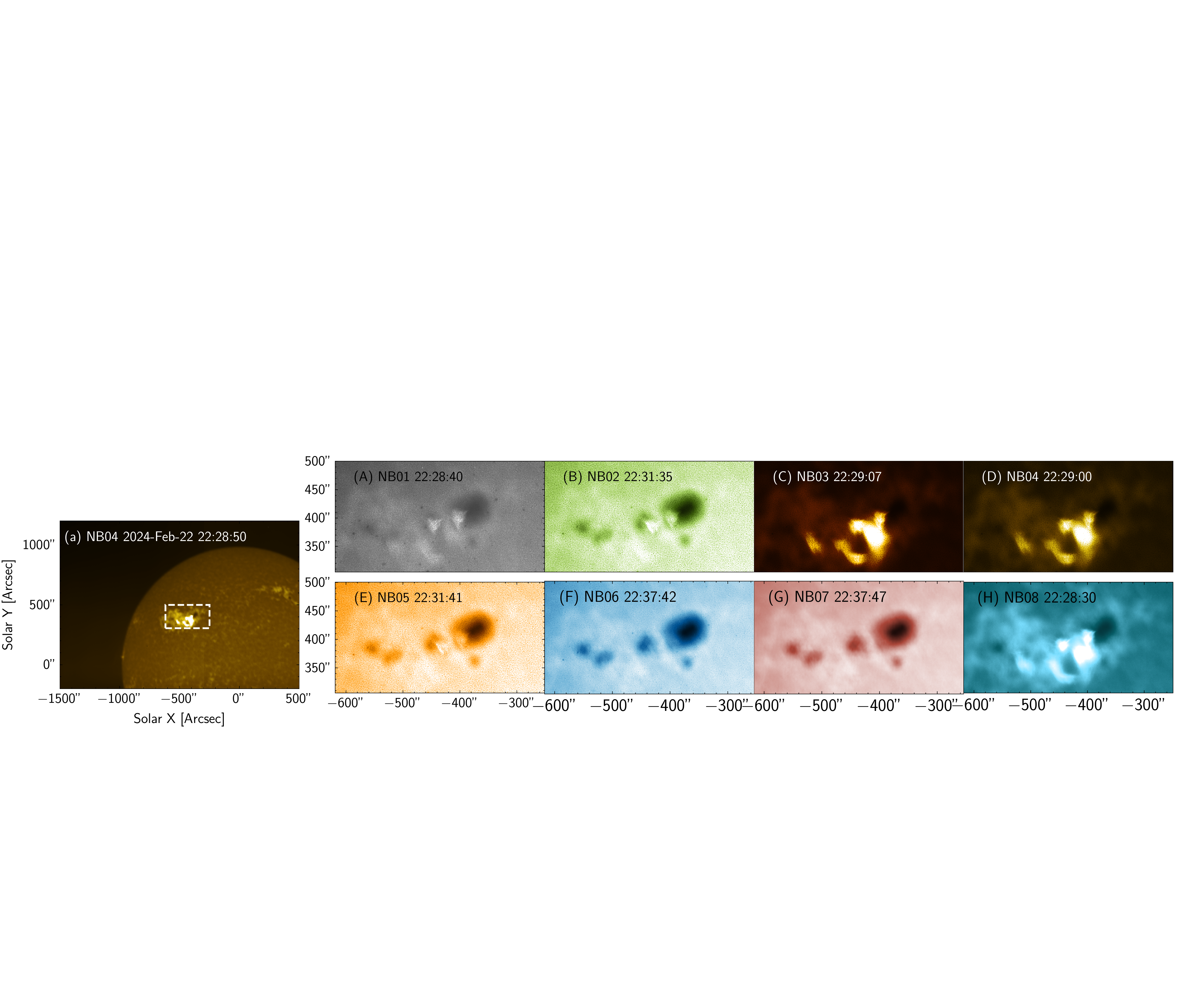}
            \caption{SUIT full-disk observation of the X6 flare on 22 February, 2024 in SUIT NB04 (Mg~\Romannum{2}~h, 280.3~nm). Panel (A{--}H) SUIT RoI observations during the peak of the respective narrow-band filters.}
            \label{fig:events3}
        \end{figure*}

        \begin{figure*}
            \centering
            \includegraphics[trim={0cm 0cm 0cm 0cm}, clip, width=0.95\linewidth]{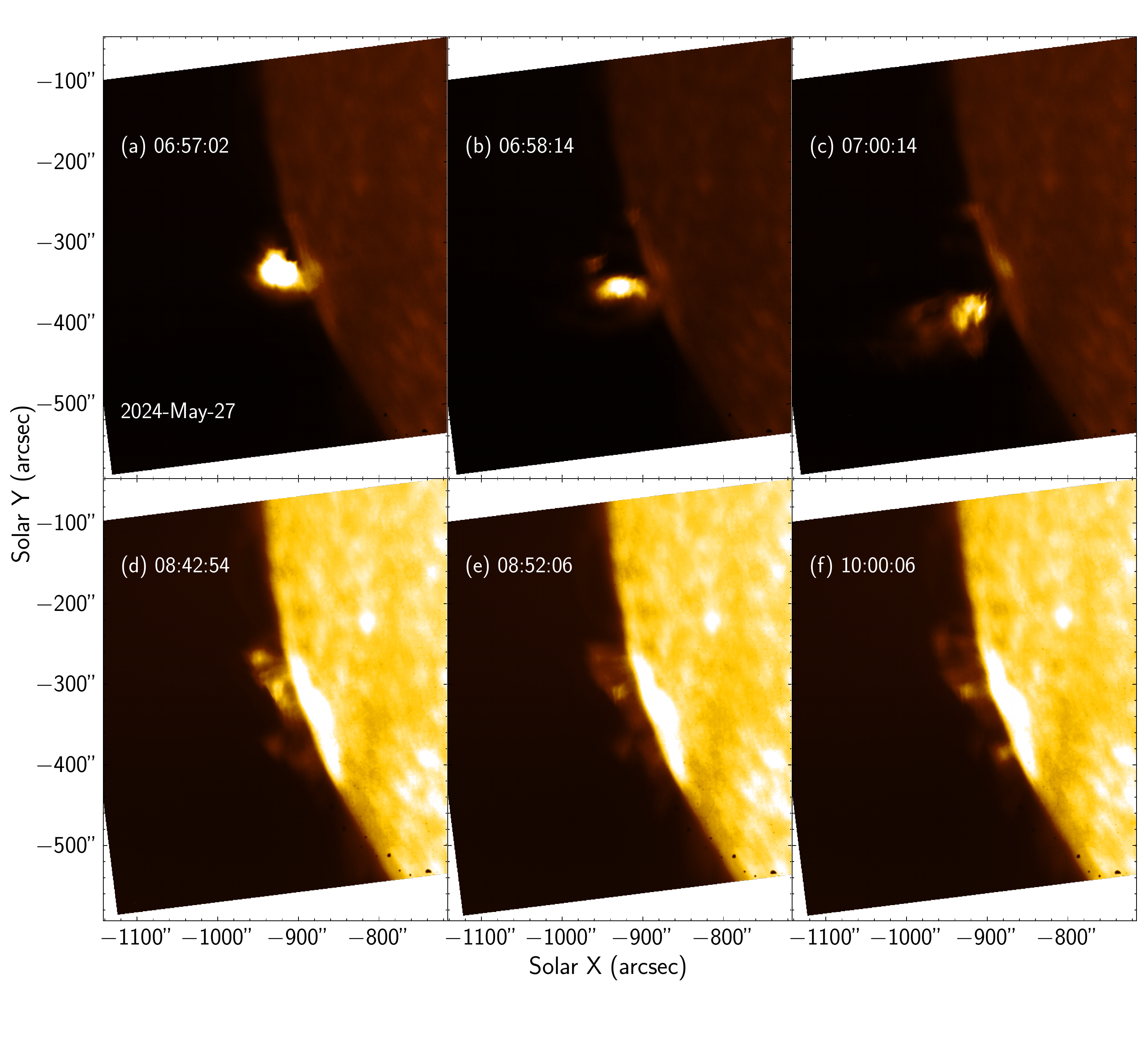}
            \caption{Sequence SUIT RoI observations in NB04 of the prominence eruption and eventual flaring loops as observed on May 27, 2024.}
            \label{fig:events4}
        \end{figure*}
        
        \item One of the final observations highlighted in this article is a prominence eruption and the associated flare observed on May 27, 2024 on the eastern limb $\approx$[$-900''$, $-350''$].
        The prominence eruption is also visible in various AIA UV and EUV channels. The large prominence eruption is followed by rising loops from behind the limbs.
        The loops are visible in Mg~\Romannum{2}~k \& h lines, Ca~\Romannum{2}~K, and the red and blue wing continuum of the Mg lines.
        The loops are also visible in H~$\mathrm{\alpha}$, Ly~$\mathrm{\alpha}$, and even the HMI white light continuum.
        The SUIT NB04 sequence of observations is shown in Fig.~\ref{fig:events4} panel (b). The flare loops are visible in Mg~\Romannum{2}~h line after $\approx$ 3 hours of prominence eruption.
    \end{itemize}

    These varied observations outline SUIT's ability to observe various solar events in the near-ultraviolet wavelength range. Various of these passbands have not been used to observe the Sun before, adding to the uniqueness of SUIT. 

\section{A nominal day of SUIT}\label{sec:nominalday}
A nominal day in SUIT observation planning begins a day before the actual observation.
During the initial phase (before the proposal driven mode starts) the SUIT team is responsible for selecting the program sequence, deciding the RoI based on the Sun's activity, and calculating the data volume resulting from all the selected sequences.
Since December 19, 2023, all proposals have been uploaded through the Aditya-L1 Proposal Processing System (PPS)\footnote{Link for Aditya-L1 PPS: \href{https://alpps.issdc.gov.in}{https://alpps.issdc.gov.in}}, an interface developed by the ISRO team.

The SUIT team operates as the payload operation center in the PPS, with the authority to upload RoIs and submit science and calibration proposals. RoIs are planned with particular interest in flaring regions, coronal holes, filaments, and quiet-Sun regions.

When the payload is open to the public, any user can submit proposals through the PPS page with the proper scientific justification and the type of observation required.
Users can specify the type of region to observe. A Time Allocation Committee (TAC) will select the proposals and decide the weekly observation plan.
The SUIT team will then determine the RoI based on user requests.
Additionally, the scientific community can request custom proposals as per their scientific requirements, including changing the position of the Sun on the SUIT CCD from the VELC-aligned position to the CCD-centered position.

The SUIT payload operation center plays a central role in this process. The data is downloaded from the satellite in strips of one-hour duration and redirected to the POC area in the Indian Space Science Data Centre (ISSDC) server.
The SUIT POC accesses the data, checks its integrity, and reports to the ISSDC team. Every morning, a data quality report is generated and sent to the ISSDC team to take necessary actions.

\section{Data products}\label{sec:data}
        \begin{figure}
            \centering
            \includegraphics[trim={5cm 6cm 6cm 2cm}, clip, width=0.95\linewidth]{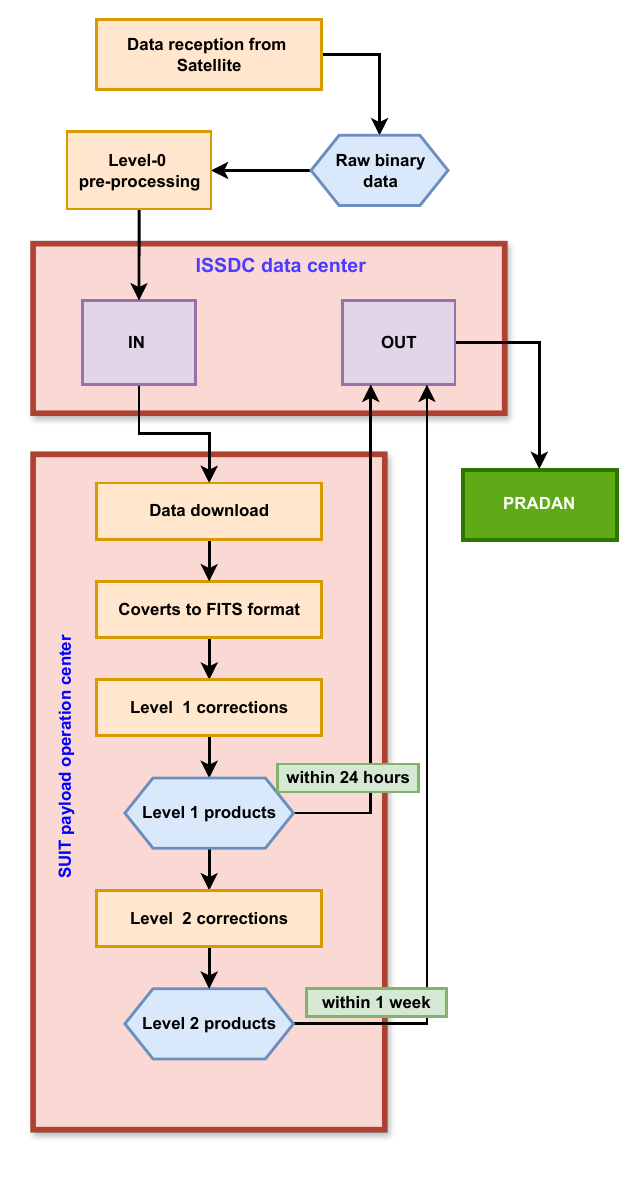}
            \caption{\revtwo{Data flow between POC, ISSDC, and PRADAN showing the exchange of raw and processed data products.}}
            \label{fig:dataflow}
        \end{figure}

The automated pipeline developed by the SUIT team pings the ISSDC server every 10 minutes to check for new data packets downloaded from the satellite.
When a new packet is found, it is transferred to the POC, and an alert is sent. The pipeline is designed in an assembly line structure, where data goes through queues, each leading to subsequent modules and queues.
This structure makes it easier for the payload team to track the processing steps.

The processing module generates level-1 science-ready data within 24 hours of data reception at the POC and sends them back to the ISSDC server.
The level-0 binary data and auxiliary files are used to generate FITS images with the necessary header information.
Further processing and image reduction are performed while generating Level 1 science-ready data products from the Level 0 FITS files.
The reduction steps include basic corrections like bias and gain, scattered light correction, and ghost correction (for the images in NB08 filter), and WCS implementation \cite{suit_test_calib}.
Level-2 data and outreach products are sent back to ISSDC for hosting within one week of data reception at the POC.
The corrections on Level-2 images include dust and spike removal, distortion correction, and PSF deconvolution \cite{suit_test_calib}.
This science-ready data is available for download by the public through the PRADAN  \revtwo{(Policy based data Retrieval, Analytics, Dissemination And Notification system)} interface, developed by the ISRO team\footnote{The SUIT data can be accessed through \href{https://pradan.issdc.gov.in/al1}{ https://pradan.issdc.gov.in/al1}}.  \revtwo{Schematic overview of the data flow between the POC, ISSDC, and the PRADAN system is illustrated in Fig~\ref{fig:dataflow}.}

\section{Conculsion}\label{sec:conclusion}
Since the opening of the door, SUIT has been continuously recording images of the Sun in the near ultraviolet, enabling us to perform studies of the solar photosphere and chromosphere.
The instrument has demonstrated stable performance, with continuous refinements in calibration and data processing, enhancing its scientific output. Several dedicated working groups have been established to address key science goals, focusing on topics including solar irradiance variations, active region dynamics, solar flares and eruptions, and quiet Sun studies.

SUIT has carried out various coordinated observations with various observatories {\it e.g.} IRIS, Solar Orbiter and Sunrise{--}\Romannum{3}. 
These coordinated efforts ensure that SUIT's data are effectively utilized to advance our understanding of solar processes. Moving forward, further optimizations in the data pipeline and operational strategies will continue to improve SUIT’s capabilities, paving the way for new discoveries in solar physics.

\section*{Acknowledgements}\label{sec:ack}
The authors acknowledge the funding support received from ISRO/DOS for the project. SUIT is built by a consortium led by the IUCAA, Pune, and supported by ISRO as part of the Aditya-L1 mission. The consortium consists of CESSI-IISER Kolkata (MoE), IIA, MAHE, MPS, USO/PRL, and Tezpur University. Aditya-L1 is an observatory class mission which is funded and operated by the ISRO. The mission was conceived and realized with the help from all ISRO Centres and payloads were realized by the payload PI Institutes in close collaboration with ISRO and many other national institutes - IIA; IUCAA; LEOS of ISRO; PRL; URSC of ISRO; VSSC of ISRO. SoLEXS is designed and developed at the Space Astronomy Group of URSC, ISRO with the help from various entities within URSC. We thank the Mission Operations and ISTRAC teams of ISRO for their support in operating SUIT payload operations. We acknowledge the use of data from the Aditya-L1, the first solar mission of the ISRO, archived at the ISSDC. We also thank the SUIT level-0 data team and the ISSDC data archive team for their invaluable support and timely assistance throughout the year. D.K. and D.T. acknowledges funding from ISRO/RESPOND for the project ``Solar Flares: Physics and Forecasting for Better Understanding of Space Weather”, ISRO/RES/2/438/22-23. We thank Professor Nandita Srivastava for her valuable comments. S.K.S. acknowledges funding from the European Research Council (ERC) under the European Union’s Horizon 2020 research and innovation programme (grant agreement No. 101097844 — project WINSUN). This SUIT pipeline used version 6.0.5 \citep{sunpy_ver} of the SunPy open source software package \citep{sunpy20} and the Python packages NumPy \citep{numpy}, Matplotlib \citep{matpltolib}, and SciencePlots \citep{SciencePlots}.
\vspace{-1em}

\bibliography{mybib}

@ARTICLE{aditya_seetha_megala_2017,
       author = {{Seetha}, S. and {Megala}, S.},
        title = "{Aditya-L1 mission}",
      journal = {Current Science},
         year = 2017,
        month = aug,
       volume = {113},
       number = {4},
        pages = {610},
          doi = {10.18520/cs/v113/i04/610-612},
       adsurl = {https://ui.adsabs.harvard.edu/abs/2017CSci..113..610S}
}

@INPROCEEDINGS{aditya,
       author = {{Tripathi}, Durgesh and {Chakrabarty}, D. and {Nandi}, A. and {Raghvendra Prasad}, B. and {Ramaprakash}, A.~N. and {Shaji}, Nigar and {Sankarasubramanian}, K. and {Satheesh Thampi}, R. and {Yadav}, V.~K.},
        title = "{The Aditya-L1 mission of ISRO}",
    booktitle = {The Era of Multi-Messenger Solar Physics},
         year = 2023,
       editor = {{Cauzzi}, Gianna and {Tritschler}, Alexandra},
       series = {IAU Symposium},
       volume = {372},
        month = jan,
        pages = {17-27},
          doi = {10.1017/S1743921323001230},
archivePrefix = {arXiv},
       eprint = {2212.13046},
 primaryClass = {astro-ph.SR},
       adsurl = {https://ui.adsabs.harvard.edu/abs/2023IAUS..372...17T}
}

@ARTICLE{suit_algo,
       author = {{Varma}, Manoj and {u'Padinhatteeri}, Sreejith and {Sinha}, Sakya and {Tyagi}, Anurag and {Burse}, Mahesh and {Yadav}, Reena and {Kumar}, Ghanshyam and {Ramaprakash}, Anamparambu and {Tripathi}, Durgesh and {Sankarasubramanian}, K. and {Nagaraju}, Krishnappa and {Vadodariya}, Koushal and {Tadepalli}, Srikar and {Deogaonkar}, Rushikesh and {Olekar}, Manjunath and {Azaruddin}, Mohamed and {Unnikrishnan}, Amrita},
        title = "{The Solar Ultra-Violet Imaging Telescope (SUIT) Onboard Intelligence for Flare Observations}",
      journal = {solphys},
         year = 2023,
        month = feb,
       volume = {298},
       number = {2},
          eid = {16},
        pages = {16},
          doi = {10.1007/s11207-023-02108-7},
       adsurl = {https://ui.adsabs.harvard.edu/abs/2023SoPh..298...16V}
}

@article{suit_main_2,
	title = {The {Solar} {Ultraviolet} {Imaging} {Telescope} on {Board} {Aditya}-{L1}},
	volume = {300},
	issn = {1573-093X},
	url = {https://doi.org/10.1007/s11207-025-02423-1},
	doi = {10.1007/s11207-025-02423-1},
	language = {en},
	number = {3},
	urldate = {2025-03-14},
	journal = {Solar Physics},
	author = {Tripathi, Durgesh and Ramaprakash, A. N. and Padinhatteeri, Sreejith and Sarkar, Janmejoy and Burse, Mahesh and Tyagi, Anurag and Kesharwani, Ravi and Sinha, Sakya and Joshi, Bhushan and Deogaonkar, Rushikesh and Roy, Soumya and Nived, V. N. and Gopalakrishnan, Rahul and Kulkarni, Akshay and Khan, Aafaque and Ghosh, Avyarthana and Rajarshi, Chaitanya and Modi, Deepa and Kumar, Ghanshyam and Yadav, Reena and Varma, Manoj and Bayanna, Raja and Chordia, Pravin and Karmakar, Mintu and Abraham, Linn and Adithya, H. N. and Adoni, Abhijit and Ahmed, Gazi A. and Banerjee, Dipankar and Bhargava Ram, B. S. and Bhandare, Rani and Chatterjee, Subhamoy and Chillal, Kalpesh and Dey, Arjun and Gandorfer, Achim and Gowda, Girish and Haridas, T. R. and Jain, Anand and James, Melvin and Jayakumar, R. P. and Justin, Evangeline Leeja and Nagaraju, K. and Kathait, Deepak and Khodade, Pravin and Kiran, Mandeep and Kohok, Abhay and Krivova, Natalie and Kumar, Nishank and Mehandiratta, Nidhi and Mestry, Vilas and Motamarri, Srikanth and Mustafa, Sajjade F. and Nandy, Dibyendu and Narendra, S. and Navle, Sonal and Parate, Nashiket and Pillai, Anju M. and Punnadi, Sujit and Rajendra, A. and Ravi, A. and Raha, Bijoy and Sankarasubramanian, K. and Sarvar, Ghulam and Shaji, Nigar and Sharma, Nidhi and Singh, Aditya and Singh, Shivam and Solanki, Sami K. and Subramanian, Vivek and T, Rethika and T, Srikanth and Thatimattala, Satyannarayana and Tota, Hari Krishna and Vishnu, T. S. and Unnikrishnan, Amrita and Vadodariya, Kaushal and Veeresha, D. R. and Venkateswaran, R.},
	month = mar,
	year = {2025},
	pages = {30},
}

@article{suit_main,
  author = {Durgesh Tripathi and A.N. Ramaprakash and Sreejith Padinhatteeri and Janmejoy Sarkar and Mahesh Burse and Ravi Kesharwani and Sakya Sinha and Bhushan Joshi and Rushikesh Deogaonkar and Soumya Roy and VN Nived and Rahul Gopalakrishnan and Akshay Kulkarni and Aafaque Khan and Avyarthana Ghosh and Deepa Modi and Anurag Tyagi and Ghanashyam Kumar and Reena Yadav and Manoj Varma and Raja Bayanna and Pravin Chordia and Mintu Karmakar and Linn Abraham and H.N. Adithya and Abhijit Adoni and Dipankar Banerjee and Rani Bhandari and Subhamoy Chatterjee and Kalpesh Chillal and Arjun Dey and Achim Gandorfer and T. R. Haridas and Melvin James and R. P. Jayakumar and Evangeline Leeja Justin and Nagaraju K and Deepak Kathait and Pravin Khodade and Mandeep Kiran and Abhay Kohok and Natalie Krivova and Nishank Kumar and Nidhi Mehandiratta and Vilas Mestry and Sreekant Motamari and Sajjade F. Mustafa and Dibyandu Nandi and S. Narendra and Sonal Navle and Nashiket Parate and Sujith Ponnadi and Chaitanya Rajarshi and A. Rajendra and Bhargava Ram and A. Ravi and Bijoy Saha and K. Sankarasubramanian and Ghulam Sarvar and Nigar Shaji and Nidhi Sharma and Aditya Singh and Shivam Singh and Sami K. Solanki and Vivek Subramanian and Rethika T and Satyannarayana Thatimattala and Hari Krishna Tota and Vishnu TS and Amrutha Unnikrishnan and Kaushal Vadodariya and D. R. Veeresha and R Venkateswaran},
  title = {The Solar Ultraviolet Imaging Telescope on board Aditya-L1},
  journal = {Solar Physics},
  year = {2025},
  status = {Accepted for publication}
}

@ARTICLE{iris,
       author = {{De Pontieu}, B. and {Title}, A.~M. and {Lemen}, J.~R. and {Kushner}, G.~D. and {Akin}, D.~J. and {Allard}, B. and {Berger}, T. and {Boerner}, P. and {Cheung}, M. and {Chou}, C. and {Drake}, J.~F. and {Duncan}, D.~W. and {Freeland}, S. and {Heyman}, G.~F. and {Hoffman}, C. and {Hurlburt}, N.~E. and {Lindgren}, R.~W. and {Mathur}, D. and {Rehse}, R. and {Sabolish}, D. and {Seguin}, R. and {Schrijver}, C.~J. and {Tarbell}, T.~D. and {W{\"u}lser}, J. -P. and {Wolfson}, C.~J. and {Yanari}, C. and {Mudge}, J. and {Nguyen-Phuc}, N. and {Timmons}, R. and {van Bezooijen}, R. and {Weingrod}, I. and {Brookner}, R. and {Butcher}, G. and {Dougherty}, B. and {Eder}, J. and {Knagenhjelm}, V. and {Larsen}, S. and {Mansir}, D. and {Phan}, L. and {Boyle}, P. and {Cheimets}, P.~N. and {DeLuca}, E.~E. and {Golub}, L. and {Gates}, R. and {Hertz}, E. and {McKillop}, S. and {Park}, S. and {Perry}, T. and {Podgorski}, W.~A. and {Reeves}, K. and {Saar}, S. and {Testa}, P. and {Tian}, H. and {Weber}, M. and {Dunn}, C. and {Eccles}, S. and {Jaeggli}, S.~A. and {Kankelborg}, C.~C. and {Mashburn}, K. and {Pust}, N. and {Springer}, L. and {Carvalho}, R. and {Kleint}, L. and {Marmie}, J. and {Mazmanian}, E. and {Pereira}, T.~M.~D. and {Sawyer}, S. and {Strong}, J. and {Worden}, S.~P. and {Carlsson}, M. and {Hansteen}, V.~H. and {Leenaarts}, J. and {Wiesmann}, M. and {Aloise}, J. and {Chu}, K. -C. and {Bush}, R.~I. and {Scherrer}, P.~H. and {Brekke}, P. and {Martinez-Sykora}, J. and {Lites}, B.~W. and {McIntosh}, S.~W. and {Uitenbroek}, H. and {Okamoto}, T.~J. and {Gummin}, M.~A. and {Auker}, G. and {Jerram}, P. and {Pool}, P. and {Waltham}, N.},
        title = "{The Interface Region Imaging Spectrograph (IRIS)}",
      journal = {solphys},
         year = 2014,
        month = jul,
       volume = {289},
       number = {7},
        pages = {2733-2779},
          doi = {10.1007/s11207-014-0485-y},
archivePrefix = {arXiv},
       eprint = {1401.2491},
 primaryClass = {astro-ph.SR},
       adsurl = {https://ui.adsabs.harvard.edu/abs/2014SoPh..289.2733D}
}

@ARTICLE{sc_filt,
       author = {{Sarkar}, Janmejoy and {Deogaonkar}, Rushikesh and {Kesharwani}, Ravi and {Padinhatteeri}, Sreejith and {Ramaprakash}, A.~N. and {Tripathi}, Durgesh and {Roy}, Soumya and {Ahmed}, Gazi A. and {Chatterjee}, Rwitika and {Ghosh}, Avyarthana and {Sankarasubramanian}, K. and {Khan}, Aafaque and {Mehandiratta}, Nidhi and {Pillai}, Netra and {Singh}, Swapnil},
        title = "{Science filter characterization of the Solar Ultraviolet Imaging Telescope (SUIT) on board Aditya-L1.}",
      journal = {Experimental Astronomy},
         year = 2025,
        month = feb,
       volume = {59},
       number = {1},
          eid = {3},
        pages = {3},
          doi = {10.1007/s10686-024-09973-5},
archivePrefix = {arXiv},
       eprint = {2412.11636},
 primaryClass = {astro-ph.SR},
       adsurl = {https://ui.adsabs.harvard.edu/abs/2025ExA....59....3S}
}

@article{dec_31st,
	title = {X-class {Flare} on 2023 {December} 31 {Observed} by the {Solar} {Ultraviolet} {Imaging} {Telescope} on {Board} {Aditya}-{L1}},
	volume = {983},
	issn = {2041-8205, 2041-8213},
	url = {https://iopscience.iop.org/article/10.3847/2041-8213/adc387},
	doi = {10.3847/2041-8213/adc387},
	number = {1},
	urldate = {2025-04-07},
	journal = {The Astrophysical Journal Letters},
	author = {Roy, Soumya and Tripathi, Durgesh and Upendran, Vishal and Padinhatteeri, Sreejith and Ramaprakash, A. N. and V. N., Nived and Sankarasubramanian, K. and Solanki, Sami K. and Sarkar, Janmejoy and Gopalakrishnan, Rahul and Deogaonkar, Rushikesh and Nandy, Dibyendu and Banerjee, Dipankar},
	month = apr,
	year = {2025},
	pages = {L6},
}

@article{feb_22nd,
	title = {Near- and {Mid}-ultraviolet {Observations} of {X}-6.3 {Flare} on 2024 {February} 22 {Recorded} by the {Solar} {Ultraviolet} {Imaging} {Telescope} on board {Aditya}-{L1}},
	volume = {981},
	issn = {2041-8205, 2041-8213},
	url = {https://iopscience.iop.org/article/10.3847/2041-8213/adb0be},
	doi = {10.3847/2041-8213/adb0be},
	number = {1},
	urldate = {2025-03-01},
	journal = {The Astrophysical Journal Letters},
	author = {Roy, Soumya and Tripathi, Durgesh and Padinhatteeri, Sreejith and Ramaprakash, A. N. and Sarwade, Abhilash R. and V. N., Nived and Sarkar, Janmejoy and Gopalakrishnan, Rahul and Deogaonkar, Rushikesh and Sankarasubramanian, K. and Solanki, Sami K. and Nandy, Dibyendu and Banerjee, Dipankar},
	month = mar,
	year = {2025},
	pages = {L19},
}

@ARTICLE{kleint17,
       author = {{Kleint}, Lucia and {Heinzel}, Petr and {Krucker}, S{\"a}m},
        title = "{On the Origin of the Flare Emission in IRIS{\textquoteright} SJI 2832 Filter:Balmer Continuum or Spectral Lines?}",
      journal = {\apj},
         year = 2017,
        month = mar,
       volume = {837},
       number = {2},
          eid = {160},
        pages = {160},
          doi = {10.3847/1538-4357/aa62fe},
archivePrefix = {arXiv},
       eprint = {1702.07167},
 primaryClass = {astro-ph.SR},
       adsurl = {https://ui.adsabs.harvard.edu/abs/2017ApJ...837..160K}
}

@ARTICLE{kowalski19,
       author = {{Kowalski}, Adam F. and {Butler}, Elizabeth and {Daw}, Adrian N. and {Fletcher}, Lyndsay and {Allred}, Joel C. and {De Pontieu}, Bart and {Kerr}, Graham S. and {Cauzzi}, Gianna},
        title = "{Spectral Evidence for Heating at Large Column Mass in Umbral Solar Flare Kernels. I. IRIS Near-UV Spectra of the X1 Solar Flare of 2014 October 25}",
      journal = {\apj},
         year = 2019,
        month = jun,
       volume = {878},
       number = {2},
          eid = {135},
        pages = {135},
          doi = {10.3847/1538-4357/ab1f8b},
archivePrefix = {arXiv},
       eprint = {1905.02111},
 primaryClass = {astro-ph.SR},
       adsurl = {https://ui.adsabs.harvard.edu/abs/2019ApJ...878..135K}
}

@ARTICLE{reetika21,
       author = {{Joshi}, Reetika and {Schmieder}, Brigitte and {Heinzel}, Petr and {Tomin}, James and {Chandra}, Ramesh and {Vilmer}, Nicole},
        title = "{Balmer continuum enhancement detected in a mini flare observed with IRIS}",
      journal = {\aap},
         year = 2021,
        month = oct,
       volume = {654},
          eid = {A31},
        pages = {A31},
          doi = {10.1051/0004-6361/202141172},
archivePrefix = {arXiv},
       eprint = {2107.11651},
 primaryClass = {astro-ph.SR},
       adsurl = {https://ui.adsabs.harvard.edu/abs/2021A&A...654A..31J}
}

@ARTICLE{heinzel14,
       author = {{Heinzel}, P. and {Kleint}, L.},
        title = "{Hydrogen Balmer Continuum in Solar Flares Detected by the Interface Region Imaging Spectrograph (IRIS)}",
      journal = {\apjl},
         year = 2014,
        month = oct,
       volume = {794},
       number = {2},
          eid = {L23},
        pages = {L23},
          doi = {10.1088/2041-8205/794/2/L23},
archivePrefix = {arXiv},
       eprint = {1409.5680},
 primaryClass = {astro-ph.SR},
       adsurl = {https://ui.adsabs.harvard.edu/abs/2014ApJ...794L..23H}
}

@ARTICLE{kleint16,
       author = {{Kleint}, Lucia and {Heinzel}, Petr and {Judge}, Phil and {Krucker}, S{\"a}m},
        title = "{Continuum Enhancements in the Ultraviolet, the Visible and the Infrared during the X1 Flare on 2014 March 29}",
      journal = {\apj},
         year = 2016,
        month = jan,
       volume = {816},
       number = {2},
          eid = {88},
        pages = {88},
          doi = {10.3847/0004-637X/816/2/88},
archivePrefix = {arXiv},
       eprint = {1511.04161},
 primaryClass = {astro-ph.SR},
       adsurl = {https://ui.adsabs.harvard.edu/abs/2016ApJ...816...88K}
}

@Article{numpy,
 title         = {Array programming with {NumPy}},
 author        = {Charles R. Harris and K. Jarrod Millman and St{\'{e}}fan J.
                 van der Walt and Ralf Gommers and Pauli Virtanen and David
                 Cournapeau and Eric Wieser and Julian Taylor and Sebastian
                 Berg and Nathaniel J. Smith and Robert Kern and Matti Picus
                 and Stephan Hoyer and Marten H. van Kerkwijk and Matthew
                 Brett and Allan Haldane and Jaime Fern{\'{a}}ndez del
                 R{\'{i}}o and Mark Wiebe and Pearu Peterson and Pierre
                 G{\'{e}}rard-Marchant and Kevin Sheppard and Tyler Reddy and
                 Warren Weckesser and Hameer Abbasi and Christoph Gohlke and
                 Travis E. Oliphant},
 year          = {2020},
 month         = sep,
 journal       = {Nature},
 volume        = {585},
 number        = {7825},
 pages         = {357--362},
 doi           = {10.1038/s41586-020-2649-2},
 publisher     = {Springer Science and Business Media {LLC}},
 url           = {https://doi.org/10.1038/s41586-020-2649-2}
}

@Article{matpltolib,
  Author    = {Hunter, J. D.},
  Title     = {Matplotlib: A 2D graphics environment},
  Journal   = {Computing in Science \& Engineering},
  Volume    = {9},
  Number    = {3},
  Pages     = {90--95},
  publisher = {IEEE COMPUTER SOC},
  doi       = {10.1109/MCSE.2007.55},
  year      = 2007
}

@ARTICLE{sunpy20,
  doi = {10.3847/1538-4357/ab4f7a},
  url = {https://iopscience.iop.org/article/10.3847/1538-4357/ab4f7a},
  author = {{The SunPy Community} and Barnes, Will T. and Bobra, Monica G. and Christe, Steven D. and Freij, Nabil and Hayes, Laura A. and Ireland, Jack and Mumford, Stuart and Perez-Suarez, David and Ryan, Daniel F. and Shih, Albert Y. and Chanda, Prateek and Glogowski, Kolja and Hewett, Russell and Hughitt, V. Keith and Hill, Andrew and Hiware, Kaustubh and Inglis, Andrew and Kirk, Michael S. F. and Konge, Sudarshan and Mason, James Paul and Maloney, Shane Anthony and Murray, Sophie A. and Panda, Asish and Park, Jongyeob and Pereira, Tiago M. D. and Reardon, Kevin and Savage, Sabrina and Sipőcz, Brigitta M. and Stansby, David and Jain, Yash and Taylor, Garrison and Yadav, Tannmay and Rajul and Dang, Trung Kien},
  title = {The SunPy Project: Open Source Development and Status of the Version 1.0 Core Package},
  journal = {The Astrophysical Journal},
  volume = {890},
  issue = {1},
  pages = {68-},
  publisher = {American Astronomical Society},
  year = {2020}
}

@software{sunpy_ver,
  author       = {Stuart J. Mumford and
                  Nabil Freij and
                  David Stansby and
                  Steven Christe and
                  Jack Ireland and
                  Florian Mayer and
                  Albert Y. Shih and
                  V. Keith Hughitt and
                  Daniel F. Ryan and
                  Simon Liedtke and
                  Laura Hayes and
                  David Pérez-Suárez and
                  Vishnunarayan K I. and
                  Will Barnes and
                  Pritish Chakraborty and
                  Andrew Inglis and
                  Punyaslok Pattnaik and
                  Brigitta Sipőcz and
                  Conor MacBride and
                  Rishabh Sharma and
                  Andrew Leonard and
                  Russell Hewett and
                  Alex Hamilton and
                  Abhijeet Manhas and
                  Asish Panda and
                  Matt Earnshaw and
                  Nitin Choudhary and
                  Ankit Kumar and
                  Raahul Singh and
                  Prateek Chanda and
                  Md Akramul Haque and
                  Michael S Kirk and
                  Michael Mueller and
                  Sudarshan Konge and
                  Rajul Srivastava and
                  Matt Wentzel-Long and
                  Yash Jain and
                  Samuel Bennett and
                  Ankit Baruah and
                  Quinn Arbolante and
                  Michael Charlton and
                  Shane Maloney and
                  Sashank Mishra and
                  Jeffrey Aaron Paul and
                  Akash Verma and
                  Nicky Chorley and
                  Aryan Chouhan and
                  Himanshu and
                  James Paul Mason and
                  Lazar Zivadinovic and
                  Sanskar Modi and
                  Yash Sharma and
                  Naman9639 and
                  Monica G. Bobra and
                  Jose Ivan Campos Rozo and
                  Larry Manley and
                  Kateryna Ivashkiv and
                  Timo Laitinen and
                  Agneet Chatterjee and
                  Johan Freiherr von Forstner and
                  Juanjo Bazán and
                  Kris Akira Stern and
                  Jan Gieseler and
                  John Evans and
                  Sarthak Jain and
                  Michael Malocha and
                  Sourav Ghosh and
                  Airmansmith97 and
                  Dominik Stańczak and
                  Rajiv Ranjan Singh and
                  Ruben De Visscher and
                  Shresth Verma and
                  SophieLemos and
                  Ankit Agrawal and
                  Arib Alam and
                  Dumindu Buddhika and
                  Himanshu Pathak and
                  Jai Ram Rideout and
                  Swapnil Sharma and
                  Jongyeob Park and
                  Matt Bates and
                  Alasdair Wilson and
                  Devansh Shukla and
                  Marius Giger and
                  Pankaj Mishra and
                  Deepankar Sharma and
                  Dhruv Goel and
                  Garrison Taylor and
                  Goran Cetusic and
                  Guntbert Reiter and
                  Jacob and
                  Mateo Inchaurrandieta and
                  Sally Dacie and
                  Sanjeev Dubey and
                  Arthur Eigenbrot and
                  Erik M. Bray and
                  Rutuja Surve and
                  Serge Zahniy and
                  Sudeep Sidhu and
                  Tomas Meszaros and
                  Utkarsh Parkhi and
                  William Russell and
                  Abhigyan Bose and
                  Abhishek Pandey and
                  Adrian Price-Whelan and
                  Amogh J and
                  André Chicrala and
                  Ankit and
                  Chloé Guennou and
                  Daniel D'Avella and
                  Daniel Williams and
                  Dipanshu Verma and
                  Jordan Ballew and
                  Krish Agrawal and
                  Nick Murphy and
                  Priyank Lodha and
                  Thomas Robitaille and
                  Tom Augspurger and
                  Yash Krishan and
                  honey and
                  neerajkulk and
                  Adwait Bhope and
                  Amarjit Singh Gaba and
                  Andrew Hill and
                  Benjamin Mampaey and
                  Bernhard M. Wiedemann and
                  Carlos Molina and
                  Daniel Garcia Briseno and
                  Duygu Keşkek and
                  Ishtyaq Habib and
                  Joseph Letts and
                  Karthikeyan Singaravelan and
                  Kritika Ranjan and
                  Noah Altunian and
                  Ole Streicher and
                  Reid Gomillion and
                  Samriddhi Agarwal and
                  Yash Kothari and
                  Yukie Nomiya and
                  mridulpandey and
                  Abigail L. Stevens and
                  Abijith B and
                  Abijith Bahuleyan and
                  Alex Kaszynski and
                  Alex W and
                  Ambar Mehrotra and
                  Andy Tang and
                  Anubhav Sinha and
                  Arfon Smith and
                  Arseniy Kustov and
                  Brandon Stone and
                  Chris Bard and
                  Emmanuel Arias and
                  Erik Tollerud and
                  Fionnlagh Mackenzie Dover and
                  Freek Verstringe and
                  Gulshan Kumar and
                  Harsh Mathur and
                  Igor Babuschkin and
                  James Calixto and
                  Jaylen Wimbish and
                  Jia Qing and
                  Juan Camilo Buitrago-Casas and
                  Kalpesh Krishna and
                  Kaustubh Chaudhari and
                  Kaustubh Hiware and
                  Koustav Ghosh and
                  MOULOUDI Mohamed Lyes and
                  Manas Mangaonkar and
                  Mark Cheung and
                  Matthew Mendero and
                  Megh Dedhia and
                  Mickaël Schoentgen and
                  Nakul Shahdadpuri and
                  Naveen Srinivasan and
                  Norbert G Gyenge and
                  Rajasekhar Reddy Mekala and
                  Ratul Das and
                  Rishabh Mishra and
                  Rohan Sharma and
                  Shashank Srikanth and
                  Shubham Jain and
                  Swapnil Kannojia and
                  Tannmay Yadav and
                  Tathagata Paul and
                  Tessa D. Wilkinson and
                  Thomas A Caswell and
                  Thomas Braccia and
                  Tiago M. D. Pereira and
                  Tim Gates and
                  Trung Kien Dang and
                  Varun Bankar and
                  William Jamieson and
                  Yudhik Agrawal and
                  platipo and
                  resakra and
                  tal66 and
                  yasintoda and
                  Raphael Attie and
                  Sophie A. Murray},
  title        = {SunPy},
  month        = apr,
  year         = 2023,
  publisher    = {Zenodo},
  version      = {v4.1.5},
  doi          = {10.5281/zenodo.7850372},
  url          = {https://doi.org/10.5281/zenodo.7850372}
}

@article{SciencePlots,
  author       = {John D. Garrett},
  title        = {{garrettj403/SciencePlots}},
  month        = sep,
  year         = 2021,
  publisher    = {Zenodo},
  version      = {1.0.9},
  doi          = {10.5281/zenodo.4106649},
  url          = {http://doi.org/10.5281/zenodo.4106649}
}

@ARTICLE{rodriguez24,
       author = {{Rodr{\'\i}guez}, Jos{\'e}-V{\'\i}ctor and {S{\'a}nchez Carrasco}, V{\'\i}ctor Manuel and {Rodr{\'\i}guez-Rodr{\'\i}guez}, Ignacio and {P{\'e}rez Aparicio}, Alejandro Jes{\'u}s and {Vaquero}, Jos{\'e} Manuel},
        title = "{Hemispheric Sunspot Number Prediction for Solar Cycles 25 and 26 Using Spectral Analysis and Machine Learning Techniques}",
      journal = {\solphys},
         year = 2024,
        month = aug,
       volume = {299},
       number = {8},
          eid = {116},
        pages = {116},
          doi = {10.1007/s11207-024-02363-2},
       adsurl = {https://ui.adsabs.harvard.edu/abs/2024SoPh..299..116R}
}

@ARTICLE{suit_test_calib,
      title={Test and Calibration of the Solar Ultraviolet Imaging Telescope (SUIT) on board Aditya-L1}, 
      author={Janmejoy Sarkar and VN Nived and Soumya Roy and Rushikesh Deogaonkar and Sreejith Padinhatteeri and Raja Bayanna and Ravi Kesharwani and A. N. Ramaprakash and Durgesh Tripathi and Rahul Gopalakrishnan and Bhushan Joshi and . Sakya Sinha and . Mahesh Burse and Manoj Varma and Anurag Tyagi and Reena Yadav and Chaitanya Rajarshi and H. N. Adithya and Abhijit Adoni and Gazi A. Ahmed and Dipankar Banerjee and Rani Bhandare and Bhargava Ram B. S. and Kalpesh Chillal and Pravin Chordia and Avyarthana Ghosh and Girish Gowda and Anand Jain and Melvin James and Evangeline Leeja Justin and Deepak Kathait and Aafaque Khan and Pravin Khodade and Abhay Kohok and Akshay Kulkarni and Ghanshyam Kumar and Nidhi Mehandiratta and Vilas Mestry and Deepa Modi and Srikanth Motamarri and K. Nagaraju and Dibyendu Nandy and S. Narendra and Sonal Navle and Nashiket Parate and Sujit Punnadi and A. Ravi and K. Sankarasubramanian and Ghulam Sarvar and Nigar Shaji and Sami K. Solanki and Rethika T and Kaushal Vadodariya and D. R. Veeresha and R Venkateswaran},
      year={2025},
      eprint={2503.23476},
      archivePrefix={arXiv},
      primaryClass={astro-ph.IM},
      url={https://arxiv.org/abs/2503.23476}, 
}

@ARTICLE{det_character,
       author = {{Varma}, S.~V. Manoj and {Tyagi}, Anurag and {Joshi}, Bhushan and {Yadav}, Reena and {Chordia}, Pravin and {Kumar}, Ghanshyam and {Sinha}, Sakya and {Burse}, Mahesh and {Padinhatteri}, Sreejith and {Deogaonkar}, Rushikesh and {Ramaprakash}, A.~N. and {Ghosh}, Avyarthana and {Tripathi}, Durgesh and {Sarkar}, Janmejoy and {Sankarasubramanian}, K. and {Nagaraju}, K. and {Vadodariya}, Koushal and {Kesharwani}, Ravi and {Khan}, Aafaque and {Olekar}, Manjunath and {Azaruddin}, Mohamed},
        title = "{The Solar Ultraviolet Imaging Telescope: detector characterization and readout electronics testing}",
      journal = {RAS Techniques and Instruments},
         year = 2023,
        month = jan,
       volume = {2},
       number = {1},
        pages = {256-263},
          doi = {10.1093/rasti/rzad013},
       adsurl = {https://ui.adsabs.harvard.edu/abs/2023RASTI...2..256V}
}

@article{pipeline_paper,
    author = {{Gopalakrishnan}, Rahul and {V. N.} Nived and et al.},
    title = "{Data Processing Pipeline of the Solar Ultraviolet Imaging Telescope (SUIT) onboard Aditya-L1}",
    note = {In Prep.},
    year = 2025,
}

@article{trace,
	title = {The transition region and coronal explorer},
	volume = {187},
	issn = {1573-093X},
	url = {https://doi.org/10.1023/A:1005166902804},
	doi = {10.1023/A:1005166902804},
	language = {en},
	number = {2},
	urldate = {2025-05-14},
	journal = {Solar Physics},
	author = {Handy, B.N. and Acton, L.W. and Kankelborg, C.C. and Wolfson, C.J. and Akin, D.J. and Bruner, M.E. and Caravalho, R. and Catura, R.C. and Chevalier, R. and Duncan, D.W. and Edwards, C.G. and Feinstein, C.N. and Freeland, S.L. and Friedlaender, F.M. and Hoffmann, C.H. and Hurlburt, N.E. and Jurcevich, B.K. and Katz, N.L. and Kelly, G.A. and Lemen, J.R. and Levay, M. and Lindgren, R.W. and Mathur, D.P. and Meyer, S.B. and Morrison, S.J. and Morrison, M.D. and Nightingale, R.W. and Pope, T.P. and Rehse, R.A. and Schrijver, C.J. and Shine, R.A. and Shing, L. and Strong, K.T. and Tarbell, T.D. and Title, A.M. and Torgerson, D.D. and Golub, L. and Bookbinder, J.A. and Caldwell, D. and Cheimets, P.N. and Davis, W.N. and Deluca, E.E. and McMullen, R.A. and Warren, H.P. and Amato, D. and Fisher, R. and Maldonado, H. and Parkinson, C.},
	month = jul,
	year = {1999},
	pages = {229--260},
}

@article{su_solar_2012,
	title = {{SOLAR} {MAGNETIZED} “{TORNADOES}:” {RELATION} {TO} {FILAMENTS}},
	volume = {756},
	issn = {2041-8205, 2041-8213},
	shorttitle = {{SOLAR} {MAGNETIZED} “{TORNADOES}},
	url = {https://iopscience.iop.org/article/10.1088/2041-8205/756/2/L41},
	doi = {10.1088/2041-8205/756/2/L41},
	number = {2},
	urldate = {2025-05-14},
	journal = {The Astrophysical Journal},
	author = {Su, Yang and Wang, Tongjiang and Veronig, Astrid and Temmer, Manuela and Gan, Weiqun},
	month = sep,
	year = {2012},
	pages = {L41},
}

@article{wedemeyer_are_2013,
	title = {{ARE} {GIANT} {TORNADOES} {THE} {LEGS} {OF} {SOLAR} {PROMINENCES}?},
	volume = {774},
	copyright = {http://iopscience.iop.org/info/page/text-and-data-mining},
	issn = {0004-637X, 1538-4357},
	url = {https://iopscience.iop.org/article/10.1088/0004-637X/774/2/123},
	doi = {10.1088/0004-637X/774/2/123},
	number = {2},
	urldate = {2025-05-14},
	journal = {The Astrophysical Journal},
	author = {Wedemeyer, Sven and Scullion, Eamon and Rouppe Van Der Voort, Luc and Bosnjak, Antonija and Antolin, Patrick},
	month = aug,
	year = {2013},
	pages = {123},
}

@inproceedings{hurlburt_heliophysics_2021,
	title = {Heliophysics {Events} {Knowledgebase} support for {Space} {Weather} {Research}},
	volume = {43},
	url = {https://ui.adsabs.harvard.edu/abs/2021cosp...43E2389H},
	urldate = {2025-06-27},
	author = {Hurlburt, Neal and Timmons, Ryan},
	month = jan,
	year = {2021},
	note = {ADS Bibcode: 2021cosp...43E2389H},
	pages = {2389},
    booktitle = {43rd COSPAR Scientific Assembly. Held 28 January - 4 February}
}

\end{document}